\documentclass{aa}  
\usepackage{graphicx}
\usepackage{txfonts}
\usepackage[dvipsnames]{xcolor}
\usepackage{siunitx} 
\usepackage{hyperref}
\hypersetup{
  colorlinks   = true, 
  urlcolor     = blue,
  linkcolor    = blue, 
  citecolor   = blue 
}

\usepackage{gensymb}
\usepackage{amssymb}
\usepackage{pifont}
\newcommand{\cmark}{\ding{51}}
\newcommand{\xmark}{\ding{55}}
\usepackage{rotating}
\usepackage{longtable}
\newcommand{\farcsec}{\mbox{\ensuremath{\hspace{1pt}.\!\!^{\prime\prime}}}}%
\begin{document} 
\titlerunning{Gas substructures in transition discs}
\authorrunning{Wölfer et al.}
   \subtitle{A survey of gas substructures as seen with ALMA}
   \title{Kinematics and brightness temperatures of transition discs}
   \author{L. W\"olfer
          \inst{1},
           S. Facchini\inst{2},
           N. van der Marel\inst{1},
           E. F. van Dishoeck\inst{1}\fnmsep\inst{3},
           M. Benisty\inst{4},
           A. J. Bohn,
           L. Francis\inst{5}\fnmsep\inst{6},\\
           A. F. Izquierdo\inst{7},
           R. D. Teague\inst{8}
          }
   \institute{Leiden Observatory, Leiden University, P.O. Box 9513, 2300 RA Leiden, The Netherlands
    \and Dipartimento di Fisica, Universit\`{a} degli Studi di Milano, Via Giovanni Celoria 16, 20133 Milano, Italy
    \and Max-Planck-Institut f\"ur extraterrestrische Physik, Gie\ss enbachstr. 1 , 85748 Garching bei M\"unchen, Germany
    \and Univ. Grenoble Alpes, CNRS, IPAG, 38000 Grenoble, France
    \and Department of Physics and Astronomy, University of Victoria, 3800 Finnerty Road, Elliot Building, Victoria, BC V8P 5C2, Canada
    \and NRC Herzberg Astronomy and Astrophysics, 5071 West Saanich Road, Victoria, BC V9E 2E7, Canada
    \and European Southern Observatory, Karl-Schwarzschild-Str. 2, 85748 Garching bei M\"unchen, Germany.
    \and Center for Astrophysics $\vert$ Harvard \& Smithsonian, 60 Garden Street, Cambridge, MA 02138, USA
   \\ e-mail: \href{mailto:woelfer@strw.leidenuniv.nl}{woelfer@strw.leidenuniv.nl}
             }
   \date{Received ; accepted}
%
%
  \abstract
   {In recent years, high-angular-resolution observations of the dust and gas content in circumstellar discs have revealed a variety of morphologies, naturally triggering the question of whether these substructures are driven by forming planets interacting with their environment or other mechanisms. While it remains difficult to directly image embedded planets, one of the most promising methods to distinguish disc-shaping mechanisms is to study the kinematics of the gas disc. Characterising deviations from Keplerian rotation can then be used to probe underlying perturbations such as planet--disc interactions. Creating spiral structures, the latter can also be traced in the brightness temperature.}
   {In this paper we aim to analyse the gas brightness temperature and kinematics of a sample of 36 transition discs observed with ALMA to resolve and characterise possible substructures that may be tracing embedded companions.}
   {For our analysis we use archival Band 6 and Band 7 ALMA observations of different CO isotopologues ($^{12}$CO, $^{13}$CO and C$^{18}$O) and fit different Keplerian disc models (thin and thick disc geomerty) to the retrieved velocity field of each disc.}
   {After subtraction of an azimuthally averaged brightness temperature profile and Keplerian rotation model from the peak brightness temperature and velocity maps, we find significant substructures in eight sources of our sample (CQ\,Tau, GG\,Tau, HD\,100453, HD\,142527, HD\,169142, HP\,Cha, TW\,Hya and UX\,Tau\,A) in both the brightness temperature and velocity residuals. Other sources show tentative features, while about half of our sample does not show any substructures in the temperature and kinematics that may be indicative of planet--disc interactions.}
   {For the first time, we compare the substructures from our analysis with various other indicators for the presence of planets. About 20\,\% of discs show strong features such as arcs or spirals, possibly associated with the presence of planets, while the majority of discs do not present as clear planet-driven signatures. Almost all discs that exhibit spirals in near-infrared scattered light show at least tentative features in the CO data. The present data are able to reveal only very massive bodies and a lack of features may suggest that, if there are planets at all, they are of lower mass (< 1-3\,$M_{\mathrm{J}}$) or may be located closer to the star within deep cavities. Deeper and higher resolution observations and modelling efforts are needed to confirm such scenarios.}
   \keywords{accretion, accretion discs --
                protoplanetary discs --
                planet-disc interactions --
                submillimeter: planetary systems 
               }
   \maketitle
%
\section{Introduction}
Circumstellar discs are formed as a consequence of angular momentum conservation during the process of star formation when material from a molecular cloud core is channeled towards the newborn star in the center. Also called protoplanetary or planet-forming discs they provide the gas and dust needed for the formation of planetary systems such as our solar system. Far from being static they evolve and eventually disperse while birthing planets, with different mechanisms shaping their appearance and the planet formation processes. At the same time, planets will interact with their environment and are expected to alter their host discs structure, leaving observable marks depending on their mass and location in the disc.

In the last decade, high-angular resolution dust and gas observations with the Atacama Large Millimeter/submillimeter Array (ALMA; \citealp{ALMA2015}) as well as near-infrared (NIR) scattered light observations with e.g the Spectro-Polarimetric High-contrast Exoplanet REsearch (SPHERE; \citealp{Sphere2019}), the Gemini
Planet Imager (GPI; \citealp{Macintosh2014}) or the Subaru telescope's High-Contrast Coronographic Imager for Adaptive Optics (HiCIAO), equipped with the Extreme Adaptive Optics System (SCExAO; \citealp{Subaro2015}) have indeed shown that a variety of substructures such as gaps or even cavities, rings, spiral arms and azimuthal asymmetries are ubiquitous in both the dust and the gas component of planet-forming discs (e.g. \citealp{Marel2013,Benisty2015,Benisty2017,Benisty2018,Casassus2016,Andrews2018,Cazzoletti2018,Feng2018,Andrews2020,Uyama2020}). 

Even though there exist several mechanisms that may explain these observations, such as gravitational instabilities (e.g. \citealp{Kratter2016}), photoevaporation (e.g. \citealp{Owen2011,Picogna2019}), magnetorotational instabilities (e.g. \citealp{Flock2015,Flock2017,Riols2019}), zonal flows (e.g. \citealp{Uribe2015}) or compositional baroclinic instabilities (e.g. \citealp{Klahr2004}), at least some of the substructures are expected to be linked to the presence of (massive) planets \citep{Lin1979,Zhang2018}. To interpret the origin of the various substructures it is crucial to understand how common they are, if they follow certain patterns and if differences/similarities can be identified for different star-disc system morphologies.

One particularly interesting subgroup of young stellar objects (YSOs) is represented by the so-called transition discs. Originally identified through a lack of infrared (IR) excess in their spectral energy distribution (SED) (\citealp{Strom1989}) they are characterised by dust (and gas) depleted inner regions (e.g. \citealp{Espaillat2014,Ercolano2017}). While they are sometimes classified as an intermediate state between a full optically thick disc and disc dispersal, planet-disc interactions provide an alternative explanation for the observed cavities. At least some transition discs -  especially those with very deep dust and gas cavities ( e.g. \citealp{Marel2016}) - are expected to be the result of dynamical clearing of a massive either planetary or binary companion. This may imply that transition discs are not an evolutionary state which every disc goes through, since massive planets (or binary companions) are not found around every star (e.g. \citealp{Johnson2010,Nielsen2019,MarelMulders2021}). Transition discs therefore represent excellent candidates to catch planet formation in action, test planet formation models and probe disc evolution mechanisms. 

To unambiguously link the observed substructures to the presence of a planet, the latter needs to be directly imaged in its environment. However, this method is only feasible for very bright and massive planets that are not severely affected by dust extinction (\citealp{Sanchis2020}). To this date, the only system in which a robust direct detection of proto-planets has been obtained is PDS 70, hosting two planets with masses of several $M_{\mathrm{J}}$ (\citealp{Keppler2018,Haffert2019,Benisty2021}). 

Alternatively we can study the indirect effects that planets may have on the dust and gas distributions. In this context, one promising method is to investigate the kinematics to look for perturbations that are induced in the velocity field of the rotating gas. Identifying deviations from Keplerian rotation can then be used to probe the local pressure gradient and to characterise the shape of the perturbation. \cite{Teague2018a} use this technique to constrain the rotation profile of HD\,163296 and its deviation from a Keplerian profile. In addition \cite{Teague2019a} report significant meridional flows in that disc. Evidence of similar meridional flows is found in HD\,169142 by \cite{Yu2021}. The kinematics of AS\,209 are studied by \cite{Teague2018b} and \cite{Rosotti2020b} who report a vertical dependence on the pressure maxima and measure the gas-dust coupling as well as the width of gas pressure bumps respectively. So-called kink-features are detected by \cite{Pinte2018a,Pinte2019} in the iso-velocity curves of HD\,163296 and HD\,97048 that are consistent with a Jupiter-mass planet. 
A possible signature of an embedded planet is also found by \cite{Casassus2019} in the HD\,100546 disc in the form of a Doppler-flip in the residual kinematics. In TW\,Hya (\citealp{Teague2019Spiral}), HD\,100453 (\citealp{Rosotti2020b}), HD\,135344B \citep{Casassus2021}, CQ\,Tau (\citealp{Woelfer2021}) and HD\,163296 \citep{Teague2021} spiral structures are found in the kinematics after subtraction of a Keplerian model, possibly connected to a companion. Non-Keplerian gas spirals are are also found in HD\,142527 by \cite{Garg2021}.
\cite{Calcino2022} show that the outer kink in HD\,163296 is possibly associated with a planetary spiral wake. \cite{Izquierdo2021a} developed a new, channel-map-fitting package to robustly identify localised velocity perturbations in both radius and azimuth and thus infer the position of an embedded planet. Applied to HD\,163296 data they are able to find indications for two embedded planets with this method \citep{Izquierdo2022}. 
\begin{table*}
    \centering
    \caption{Stellar properties, outer disc inclination and dust cavity radius of the disc sample studied in this work.}
    \begin{tabular}{lccccccccc}
    \hline
    \hline
        Object & $d$ (pc) & Spectral Type & $T_{\mathrm{eff}}$ (K) & $L$ ($L_{\odot}$) & $M_*$ ($M_{\odot}$) & i ($\degree$) & Classification & Dust Cavity (au) & Ref.\tablefootmark{a} \\
        \hline
        AA\,Tau & 137 & K7 & 4350 & 1.1 & 0.68 & 59 & TTS & 44 & 1\\
        AB\,Aur & 163 & A0 & 9520 & 65.1 & 2.56 & 23 & Herbig & 156 & 1\\
        CQ\,Tau & 163 & F2 & 6890 & 10.0 & 1.63 & 35 & Herbig & 50 & 1\\
        CS\,Cha & 176 & K2 & 4780 & 1.9 & 1.4 & 8 & TTS & 37 & 1\\
        DM\,Tau & 145 & M2 & 3580 & 0.2 & 0.39 & 35 & TTS & 25 & 1\\
        DoAr\,44 & 146 & K2 & 4780 & 1.9 & 1.4 & 20 & TTS & 40 & 1\\
        GG\,Tau & 140 & K7+M0 & 4060 & 1.6 & 0.66 & 36 & TTS & 224 & 1\\
        GM\,Aur & 160 & K5 & 4350 & 1.0 & 1.01 & 53 & TTS & 40 & 1\\
        HD\,100453 & 104 & F0 & 7200 & 6.2 & 1.47 & 30 & Herbig & 30 & 1\\
        HD\,100546 & 110 & A0 & 9520 & 25.0 & 2.13 & 42 & Herbig & 27 & 1\\
        HD\,135344B & 136 & F5 & 6440 & 6.7 & 1.51 & 12 & Herbig & 52 & 1\\
        HD\,139614 & 134 & A9 & 7750 & 6.0 & 1.57 & 18 & Herbig & - & 2 \\
        HD\,142527 & 157 & F6+M5/M6 & 6360 & 9.9 & 1.69+0.26 & 27 & Herbig & 185 & 1,3\\
        HD\,169142 & 114 & A5 & 8200 & 8.0 & 1.65 & 12 & Herbig & 26 & 1\\
        HD\,34282 & 312 & A0 & 9520 & 10.8 & 2.11 & 59 & Herbig & 87 & 1\\
        HD\,97048 & 185 & A0 & 9520 & 30.0 & 2.17 & 41 & Herbig & 63 & 1\\
        HP\,Cha & 160 & K7 & 4060 & 2.4 & 0.95 & 37 & TTS & 50 & 1\\
        IP\,Tau & 131 & M0 & 3850 & 0.6 & 0.54 & 45 & TTS & 25 & 1\\
        IRS\,48 & 134 & A0 & 9520 & 17.8 & 1.96 & 50 & Herbig & 83 & 1\\
        J1604.3-2130 & 150 & K3 & 4780 & 0.7 & 1.1 & 6 & TTS & 87 & 1\\ %
        LkCa\,15 & 159 & K2 & 4730 & 1.3 & 1.32 & 55 & TTS & 76 & 1\\
        MWC\,758 & 160 & A7 & 7850 & 14.0 & 1.77 & 21 & Herbig & 62 & 1\\
        PDS\,70 & 113 & K7 & 4060 & 0.3 & 0.8 & 52 & TTS & 74 & 1\\
        PDS\,99 & 155 & K6 & 4205 & 1.1 & 0.88 & 55 & TTS & 56 & 1\\
        RXJ1615.3-3255 & 156 & K7 & 4100 & 0.6 & 0.73 & 47 & TTS & - & 2 \\ 
        RXJ1852.3-3700 & 146 & K2 & 4780 & 0.6 & 1.05 & 30 & TTS & 49 & 1\\
        RY\,Lup & 159 & K2 & 4780 & 1.9 & 1.4 & 67 & TTS & 69 & 1\\
        RY\,Tau & 175 & G2 & 5860 & 15.0 & 2.25 & 65 & TTS & 27 & 1\\
        SR\,21 & 138 & G4 & 5770 & 11.0 & 2.12 & 16 & TTS & 56 & 1\\
        Sz\,91 & 159 & M1 & 3850 & 0.2 & 0.54 & 45 & TTS & 86 & 1\\
        SZ\,Cha & 190 & K2 & 5100 & 1.7 & 1.45 & 47 & TTS & - & 2\\
        T\,Cha & 107 & G8 & 5570 & 1.3 & 1.12 & 73 & TTS & 34 & 1\\
        TW\,Hya & 60 & K7 & 4205 & 0.3 & 0.81 & 7 & TTS & 2 & 1\\
        UX\,Tau\,A & 140 & G8 & 5570 & 2.5 & 1.4 & 40 & TTS & 31 & 1\\
        V1247\,Ori & 400 & F0 & 7200 & 15.0 & 1.82 & 30 & Herbig & 64 & 1\\
        V4046\,Sgr & 72 & K7+K5 & 4060 & 0.5 & 0.9+0.85 & 34 & TTS & 31 & 1,4 \\
    \hline 
    \end{tabular}
    \tablefoot{
    \tablefoottext{a}{Unless indicated otherwise, data for spectral type, distance, effective temperature, stellar luminosity, stellar mass and disc inclination are taken from (1) \cite{Francis2020} where all original references can be found. The distances are according to \cite{Gaia2018}. (2) \cite{Bohn2022}, (3) \cite{Claudi2019}, (4) \cite{Rosenfeld2012}}. The radius of the dust cavity was determined by \cite{Francis2020}. 
        }
    \label{tab:stellarProp}
\end{table*}

Besides the kinematics, it can also be useful to look for substructures or asymmetries in the peak intensity/brightness temperature residuals in the search for evidence of companions. The density waves created by a companion will result in an increase in surface density and thus in a higher opacity. This will move the $\tau = 1$ layer to a higher altitude where the temperature is generally higher, resulting in spiral substructures in the gas brightness temperature (\citealp{Phuong2020b,Phuong2020a}). In addition, planets can generate tightly wound spirals in the brighness temperature through buoyancy resonances \citep{Bae2021}. The temperature structure in planet-driven spiral arms is investigated by \cite{Muley2021} and their models may explain the observed thermal features in discs like TW\,Hya and CQ\,Tau: \cite{Teague2019Spiral} and \cite{Woelfer2021} report the detection of spiral structures in the $^{12}$CO brightness temperature of TW\,Hya and CQ\,Tau respectively after subtraction of an azimuthally averaged model. These spirals are (at least partly) linked to the spirals observed in the velocity residuals (for TW\,Hya see also \cite{Sturm2020}), and in the case of CQ\,Tau connected to a small spiral in the NIR (\citealp{Uyama2020}). 

Studying the gas component in discs may enable to assess the different dynamical processes described above that are shaping the disc and reveal previously undetected substructures. In this context, probing different disc layers with various molecules may help distinguish the formation mechanisms of the observed substructures (e.g \citealp{Pinte2018, Law2021}). For example, in a passively heated disc with a positive vertical temperature gradient, more tightly wound spirals are expected towards the midplane in the planetary scenario, while similar spiral pitch angles would be established between the surface and midplane layers if resulting from gravitational instabilities (\citealp{Juhasz2018}). Furthermore, an embedded planet will induce perturbations in all three velocity components, which will vary as a function of height: The magnitude of radial and rotational perturbations ($v_{\mathrm{r}}$, $v_{\mathrm{\varphi}}$) decreases and that of vertical perturbations ($v_{\mathrm{z}}$) increases towards the disc surface \citep{Pinte2019}.  

To this point, the connection between inner and outer disc structures in protoplanetary discs is not fully understood but represents an important piece in the planet formation puzzle. Several observations of transition discs in NIR scattered light have revealed dark regions (e.g. \citealp{Stolker2016, Casassus2018}) which are commonly interpreted as shadows resulting from a misalignment between the inner and the outer disc (e.g. \citealp{Marino2015, Facchini2018}). One, particularly exciting explanation for this is the presence of (one or several) massive misaligned companions that induce a misalignment in specific disc regions around them \citep{Francis2020,Perraut2021,Bohn2022}.      

In this work, we investigate archival CO data of a sample of 36 transition discs in terms of both their velocity and brightness temperature structure to search for possible perturbations and features that may be linked to the presence of embedded companions. The paper is structured as follows: In \hyperref[sec:observations]{Sect.~\ref*{sec:observations}} we give an overview of the selected targets. The observational results, including brightness temperature and velocity maps, as well as radial intensity profiles are presented in \hyperref[sec:results]{Sect.~\ref*{sec:results}}. In \hyperref[sec:analysis]{Sect.~\ref*{sec:analysis}} we describe our analysis, showing the resulting velocity and brightness temperature residuals. These results are discussed in \hyperref[sec:Discussion]{Sect.~\ref*{sec:Discussion}} where a comparison with other indicators of planets is done. A summary of our work is presented in \hyperref[sec:Summary]{Sect.~\ref*{sec:Summary}}. 
\begin{table*}
    \centering
    \caption{Characteristics of the ALMA line data for the main lines of this analysis.}
    \begin{tabular}{lcccccccc}
    \hline
    \hline
        Object & Line & ALMA Project ID & Beam ($\arcsec$) & LAS ($\arcsec$) & $\Delta \upsilon$ (km$\,$s$^{-1}$) & RMS (mJy$\,$beam$^{-1}$) & Cube Source\tablefootmark{a} \\
        \hline
        AA\,Tau & $^{13}$CO 3-2 & 2015.1.01017.S & 0.28x0.22 & 8.0 & 0.11 & 13.5 & A \\
        AB\,Aur & $^{13}$CO 3-2 & 2012.1.00303.S & 0.37x0.23 & 7.2 & 0.2 & 6.2 & P/PC & \\ 
        CQ\,Tau & $^{12}$CO 2-1 & 2013.1.00498.S & 0.12x0.1 & 5.3 & 0.5 & 1.2 & P/PC\\
        & & 2016.A.00026.S & & 2.9\\
        & & 2017.1.01404.S & & 2.7\\
        CS\,Cha & $^{12}$CO 3-2 & 2017.1.00969.S & 0.1x0.07 & 2.4 & 0.11 & 4.2 & A &  \\
        DM\,Tau & $^{12}$CO 2-1 & 2016.1.00724.S & 0.86x0.8 & 10.5 & 0.08 & 17.6 & A & \\
        DoAr\,44 & $^{13}$CO 3-2 & 2012.1.00158.S & 0.25x0.19 & 3.2 & 0.5 & 13.7 & P/PC & \\
        GG\,Tau & $^{12}$CO 2-1 & 2018.1.00532.S & 0.34x0.27 & 9.7 & 0.08 & 2.6 & A & \\
        GM\,Aur & $^{12}$CO 2-1 & 2018.1.01055.L & 0.15x0.15 & 3.6-44.1 & 0.2 & 2.8 & P/PC\\
        HD\,100453 & $^{12}$CO 3-2 & 2017.1.01424.S & 0.05x0.05 & 1.3 & 0.42 & 1.0 & P/PC & \\
        HD\,100546 & $^{12}$CO 2-1 & 2016.1.00344.S & 0.08x0.06 & 1.1/2.7 & 0.5 & 1.2 & P/PC & \\ 
        HD\,135344B & $^{13}$CO 3-2 & 2012.1.00158.S & 0.26x0.21 & 3.1 & 0.24 & 19.1 & P/PC & \\
        HD\,139614 & $^{13}$CO 2-1 & 2015.1.01600.S & 0.77x0.55 & 8.4 & 0.4 & 26.5 & P/PC\\
        HD\,142527 & $^{12}$CO 2-1 & 2015.1.01353.S & 0.28x0.26 & 3.9 & 0.09 & 2.5 & A\\
        HD\,169142 & $^{12}$CO 2-1 & 2015.1.00490.S & 0.18x0.13 & 4.2 & 0.05 & 1.2 & P/PC\\
        HD\,34282 & $^{12}$CO 3-2 & 2013.1.00658.S & 0.26x0.2 & 5.1/9.7 & 0.2 & 8.2 & P/PC\\
        HD\,97048 & $^{13}$CO 3-2 & 2016.1.00826.S & 0.11x0.07 & 1.8/4.4 & 0.12 & 3.8 & P/PC\\
        HP\,Cha & $^{12}$CO 2-1 & 2016.1.00583.S & 0.3x0.21 & 6.2 & 0.63 & 3.7 & A\\
        IP\,Tau & $^{12}$CO 2-1 & 2013.1.00163.S & 0.24x0.21 & 3.2 & 1.0 & 5.5 & P/PC \\
        IRS\,48 & $^{12}$CO 3-2 & 2013.1.00100.S & 0.19x0.13 & 1.4-3.5 & 0.24 & 8.3 & P/PC \\
        J1604 & $^{12}$CO 3-2 & 2015.1.00888.S & 0.23x0.19 & 3.3/6.7 & 0.21 & 7.1 & P/PC\\ 
        LkCa\,15 & $^{12}$CO 2-1 & 2018.1.01255.S & 0.41x0.29 & 7.6 & 0.04 & 6.0 & P/PC \\
        MWC\,758 & $^{13}$CO 3-2 & 2012.1.00725.S & 0.19x0.16 & 5.0 & 0.11 & 8.9 & P/PC \\
        PDS\,70 & $^{12}$CO 3-2 & 2017.A.00006.S & 0.11x0.1 & 2.3 & 0.42 & 1.1 & P/PC \\
        PDS\,99 & $^{12}$CO 2-1 & 2015.1.01301.S & 0.3x0.22 & 4.9 & 0.16 & 8.3 & A \\
        RXJ1615 & $^{12}$CO 3-2 & 2012.1.00870.S & 0.3x0.23 & 3.2/6.5 & 0.21 & 14.1 & P/PC\\
        RXJ1852 & $^{12}$CO 2-1 & 2018.1.00689.S & 0.16x0.12 & 1.8 & 0.63 & 4.3 & A \\
        RY\,Lup & $^{12}$CO 3-2 & 2017.1.00449.S & 0.22x0.17 & 2.71 & 0.85 & 4.0 & P/PC\\
        RY\,Tau & $^{12}$CO 2-1 & 2013.1.00498.S & 0.28x0.16 & 1.67 & 0.5 & 9.1 & A\\
        SR\,21 & $^{12}$CO 2-1 & 2018.1.00689.S & 0.14x0.12 & 1.71 & 0.64 & 4.8 & A\\
        Sz\,91 & $^{12}$CO 3-2 & 2012.1.00761.S & 0.17x0.13 & 1.36 & 0.2 & 11.8 & A \\
        SZ\,Cha & $^{12}$CO 3-2 & 2013.1.01075.S & 0.82x0.43 & 3.70 & 0.5 & 26.2 & P/PC\\
        T\,Cha & $^{12}$CO 2-1 & 2017.1.01419.S & 0.24x0.17 & 2.55 & 0.32 & 9.0 & A\\
        TW\,Hya & $^{12}$CO 3-2 & 2015.1.00686.S & 0.14x0.13 & 0.37 & 0.25 & 3.5 & P/PC\\
        & & 2016.1.00629.S & & 1.3/6.0\\
        UX\,Tau\,A & $^{12}$CO 3-2 & 2015.1.00888.S & 0.2x0.16 & 2.41 & 0.21 & 3.4 & P/PC\\
        V1247\,Ori & $^{12}$CO 3-2 & 2016.1.01344.S & 0.05x0.03 & 0.86 & 1.0 & 2.0 & P/PC\\
        V4046\,Sgr & $^{12}$CO 2-1 & 2016.1.00724.S & 0.41x0.29 & 4.48 & 0.08 & 9.9 & A\\
    \hline
    \end{tabular}
    \tablefoot{
        \tablefoottext{a}{P/PC: Reimaged data cube. Public data or obtained via private communication, A: Archival data product.}
        }
    \label{tab:dataProp}
\end{table*}
\begin{figure*}[h!]
\centering
\includegraphics[width=1.0\textwidth]{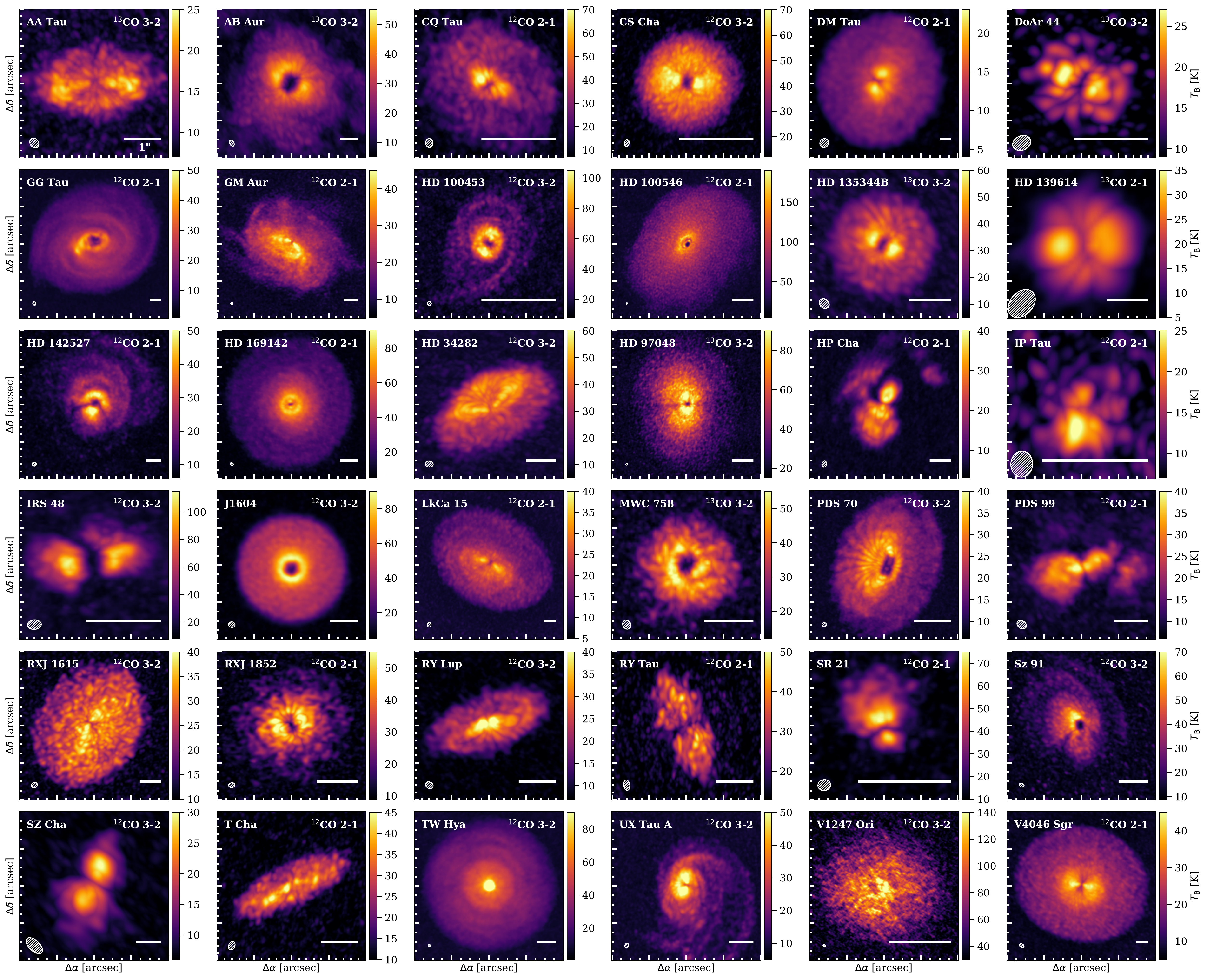}
\caption{Peak brightness temperature maps of the gas emission in our targets, shown for the main CO lines used in this analysis. The conversion from peak intensity to units of Kelvin is done with the Planck law. The circle and bar in the bottom left and bottom right corner of each panel indicate the beam and a $1\arcsec$ scale respectively. }\label{fig:TbMapsMajor}
\end{figure*}
\section{Observations}\label{sec:observations}
Our selected sample consists of 36 transition discs, chosen from the sample of \cite{Francis2020} where sufficient CO data are available. Except for TW\,Hya, these discs show large (> 25\,au) inner dust cavities and therefore represent the ideal candidates to search for planet-disc interactions. It comprises different star-disc system architectures, including a range of spectral types (M2 to A0; primary) and stellar masses (0.4\,$M_{\odot}$ to 2.6\,$M_{\odot}$, primary) counting 23 TTauri and 13 Herbig stars. Some stellar and disc properties of our targets are listed in \autoref{tab:stellarProp}.

For our analysis we collect either Band 6 or Band 7 archival CO line data, observed with ALMA. For most sources (two third) we use reimaged/self-calibrated data cubes that are either public or were obtained via private communication, for the remaining sources we use archival data products. This is indicated in \autoref{tab:dataProp}, where some characteristics of the data cubes are listed for the main lines used in our study. Typical spectral resolutions of the data are a few 100 m\,s$^{-1}$ and spatial resolutions lie between $\sim 6 -135$\,au (median: 31\,au). RMS values lie between $\sim$0.5-47\,K (median: 3.4\,K) when scaled for a channel width of 100\,m\,s$^{-1}$. 

To assess if combining reimaged and archival products affects our results we compare the reimaged data sets with the archival products for the same data set. We find that the Keplerian fit (see \hyperref[sec:velocityRes]{Sect.~\ref*{sec:velocityRes}}) is not significantly affected. The detection of extended (over several beams) substructures such as spirals or the non-detection of features is also not affected. Only tentative features are sometimes only visible in the reimaged data. Some examples for this test are shown in \hyperref[fig:compClean]{Fig.~\ref*{fig:compClean}} in the Appendix: While clear spirals are found in both the reimaged and archival product data of UX\,Tau\,A, a tentative spiral in the brightness temperature of HD\,135344B and spiral/arc in the kinematics of J1604 are only visible in the reimaged data. RXJ\,1615 on the other hand shows no clear spirals or arcs in both data products. 

Several discs in our sample are affected by cloud absorption, namely AB\,Aur, HD\,142527, HD\,97048, HP\,Cha, IP\,Tau, IRS\,48, PDS\,99, SR\,21, Sz\,91 and SZ\,Cha. We mask the regions affected by this in the calculations of radial profiles (\hyperref[sec:radprof]{Sect.~\ref*{sec:radprof}}) and brightness temperature residuals (\hyperref[sec:tempres]{Sect.~\ref*{sec:tempres}}). For some of the targets we analysed additional CO isotopologues, that are listed with the data properties in the Appendix in \autoref{tab:dataPropAdd}. Our main lines for analysis (\autoref{tab:dataProp}) are based on their brightness as well as the spatial and spectral resolution of the observation. A few of our targets have already been analysed with the same techniques, as explained in \hyperref[sec:analysis]{Sect.~\ref*{sec:analysis}} (CQ\,Tau \citep{Woelfer2021}, HD\,100453 \citep{Rosotti2020a}, TW\,Hya \citep{Teague2019Spiral}), but are included in this work for comparison.  
\section{Observational results}\label{sec:results}
\subsection{Brightness temperature}
In \hyperref[fig:TbMapsMajor]{Fig.~\ref*{fig:TbMapsMajor}} we present the peak brightness temperature maps (continuum subtracted) for the main CO lines (mostly $^{12}$CO, some $^{13}$CO). Maps for the additional lines can be found in \hyperref[fig:TbAddBand6]{Fig.~\ref*{fig:TbAddBand6}} and \hyperref[fig:TbAddBand7]{Fig.~\ref*{fig:TbAddBand7}} in the Appendix. The maps presented in \hyperref[fig:TbMapsMajor]{Fig.~\ref*{fig:TbMapsMajor}} are shown again in \hyperref[fig:TbMapsMajorCont]{Fig.~\ref*{fig:TbMapsMajorCont}} with overlaid continuum images, illustrating that for most targets the dust disc (mm-sized grains, B6 and B7) is substantially smaller than the gas disc. This can readily be explained by radial drift processes and/or by the difference between dust and gas opacities \citep{Facchini2017,Trapman2019}. 

To compute the peak intensity maps we use the standard moment 8 implementation in the \texttt{bettermoments} code \citep{bettermoments} and then convert from flux density units to units of Kelvin with the Planck law. No masking is applied in the computation of the maps. The brightness temperature traces a combination of kinetic gas temperature and column density, with the optically thick lines mostly measuring the temperature while the more optically thin lines mostly trace the column density. In this context, the observed gas temperatures of $\sim 30$\,K up to $\sim 200$\,K are as expected in the upper disc layers (see e.g.  \citealp{Bruderer2013,Bruderer2014, Leemker2022}). 
As discussed below, some of the discs show interesting features in their peak brightness temperature. Very massive companions are clearly observable from the peak brightness temperature by their ability to induce spirals which are prominent enough to be observed with this data quality. Clear spiral structures can be discerned in GG\,Tau, HD\,100453, HP\,Cha and UX\,Tau\,A. GG\,Tau is (at least) a quadruple star system surrounded by a massive disc, with the substructures likely tracing star-disc or planet-disc interactions \citep{Leinert1991,Dutrey2014,Phuong2020b}. HD\,100453 and UX\,Tau\,A are also known to have stellar companions that are responsible for the observed spirals \citep{Rosotti2020a,menard2020}. HP\,Cha is affected by cloud absorption on the blue shifted side, but the extended (red-shifted) structure suggests interactions with the environment such as infalling material from a streamer or a fly-by. Even though the disc around V4046\,Sgr is also known to be circumbinary, no clear substructures can be seen. The reason for this is that the two stars in the system are orbiting each other at a small distance ($<1\,\mathrm{au}$, $2.4\,\mathrm{d}$, \citealp{Stempels2004}), acting like a single gravitational point source on much larger scales.
\begin{figure*}
\centering
\includegraphics[width=1.0\textwidth]{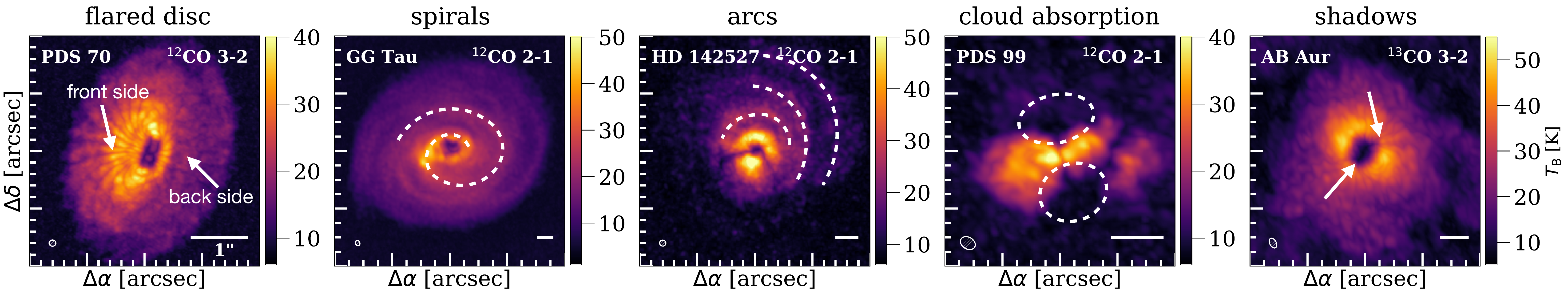}
\caption{Examples of the different features seen in the brightness temperature of our sources. The circle and bar in the bottom left and bottom right corner of each panel indicate the beam and a $1\arcsec$ scale respectively.}\label{fig:examplesFeatures}
\end{figure*}

Indications of spirals are visible in CQ\,Tau, where one side of the disc is substantially brighter, representing the anchoring point of the spiral (see \citealp{Woelfer2021}) as well as in HD\,135344B and TW\,Hya. For HD\,135344B similar spiral features have been found by \cite{Casassus2021} in $^{12}$CO 2-1 data. Other discs show arc-like azimuthal asymmetries like HD\,142527 (in all CO isotopologues, \citealp{Casassus2015,Garg2021,Yen2020}) or Sz\,91. \cite{Tsukagoshi2019} explain the arc-like structure in Sz\,91 with a flared disc, showing emission from the front and the back side, in combination with a dust ring.
\begin{figure*}
\centering
\includegraphics[width=1.0\textwidth]{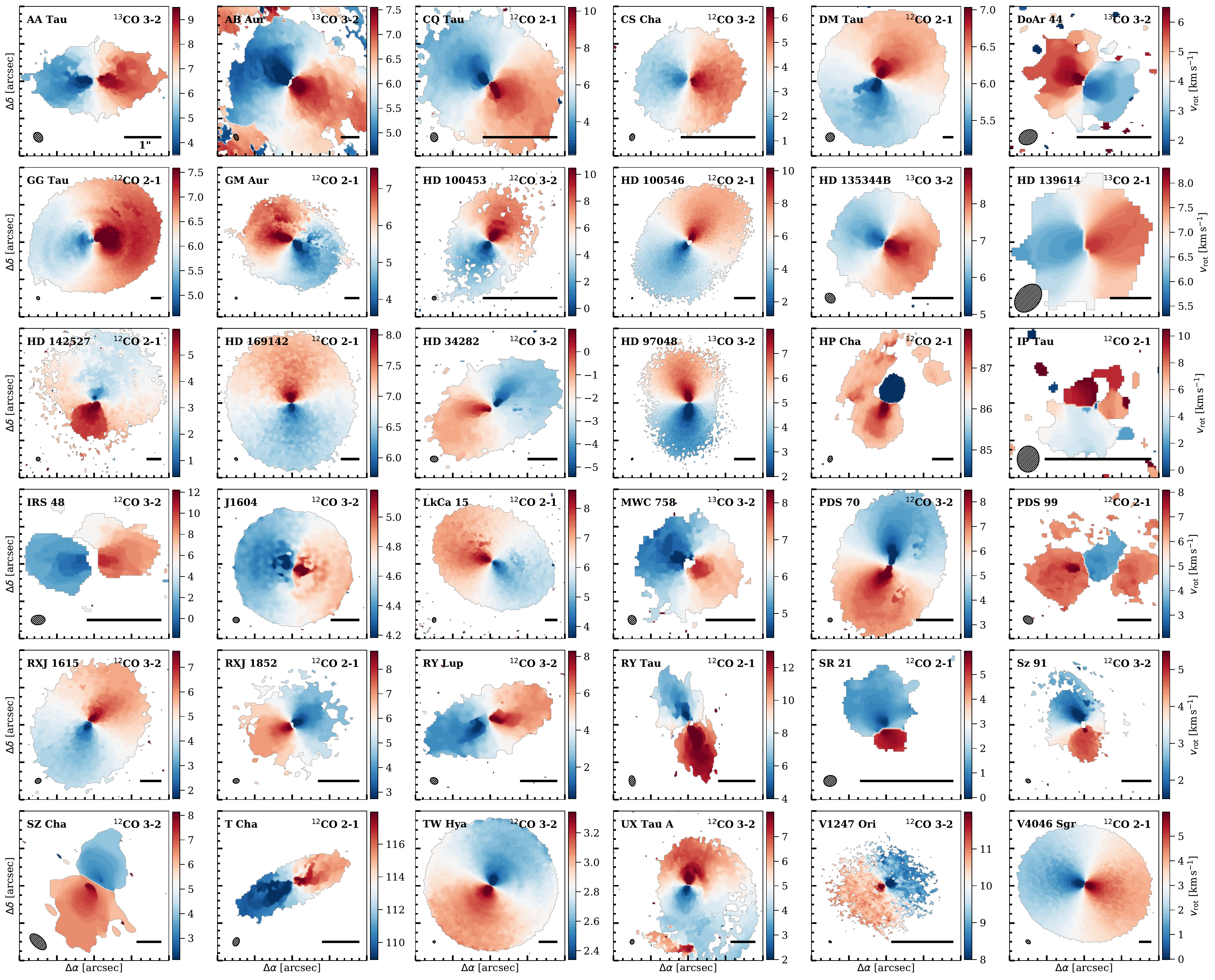}
\caption{Rotation velocity maps of the gas emission in our targets, shown for the main CO lines used in this analysis. The maps are computed with \texttt{bettermoments}. The circle and bar in the bottom left and bottom right corner of each panel indicate the beam and a $1\arcsec$ scale respectively.}\label{fig:vrotMapsMajor}
\end{figure*}

In a couple of maps - e.g of AB\,Aur, CQ\,Tau or MWC\,758 - symmetric dimmed regions are visible. Such features are commonly linked to the presence of a misaligned inner disc, casting a shadow over the outer disc (e.g. \citealp{Marino2015, Facchini2018}). Beam dilution effects can however cause artificial features along the minor axis (see example of CQ Tau in \citealp{Woelfer2021}), thus caution should be taken when interpreting these under-brightnesses. The misalignment hypothesis has recently been tested by \cite{Bohn2022} for a sub-sample of our discs by comparing position angle and inclination of the inner disc measured with VLTI/GRAVITY and the outer disc measured with ALMA. Significant misalignments are found for CQ\,Tau, HD\,100453, HD\,142527, HD\,34282, RY\,Lup and V1247\,Ori. \cite{Francis2020} also find misalignments from ALMA inner disc images, which are significant for eight sources in either position angle or inclination (AB\,Aur, GG\,Tau, HP\,Cha, MWC\,758, PDS\,70, SR\,24\,S, TW\,Hya, V4046\,Sgr). In \hyperref[fig:examplesFeatures]{Fig.~\ref*{fig:examplesFeatures}}, some examples are given for the different features that can be observed in the brightness temperature. We note that arcs can also be seen as part of a spiral. In this work we identify spirals as structures covering a larger range of radii, while arcs are mostly observed at one radius.   

To un-cover small substructures in the brightness temperature structure, we further analyse these maps in \hyperref[sec:tempres]{Sect.~\ref*{sec:tempres}} by subtracting azimuthally symmetric brightness temperature profiles from the data. 
\subsection{Rotation velocity}
In \hyperref[fig:vrotMapsMajor]{Fig.~\ref*{fig:vrotMapsMajor}} we present the kinematics of our targets, again showing the main lines, while the additional lines can be found in the Appendix in \hyperref[fig:v0AddBand6]{Fig.~\ref*{fig:v0AddBand6}} and \hyperref[fig:v0AddBand6]{Fig.~\ref*{fig:v0AddBand7}}. To compute the line-of-sight velocity of the gas we use the quadratic method implemented in the \texttt{bettermoments} package: a quadratic function is fitted to the brightest pixel in the spectrum as well as the two neighbouring pixels to find the centroid of the line in pixel coordinates. To reduce the noise at the disc edge, we apply a masking for regions below a certain signal-to-noise ratio (S/N). The magnitude of this clipping is obtained via inspection of each individual map, ranging between 2\,$\sigma$ to 5\,$\sigma$. 

Even though the spiral features are not as prominent in the kinematics as in brightness temperature maps, they are still observable in GG\,Tau, HD\,100453, HP\,Cha and UX\,Tau\,A. CQ\,Tau, HD\,135344B and TW\,Hya show indications of spirals in the brightness temperature but in the kinematics these indications are only present for TW\,Hya. CQ\,Tau however shows twisted kinematics in the center that resemble a warp but are likely caused by the spiral structure \citep{Woelfer2021}. Clearly twisted kinematics can also be seen in the center of HD\,142527, for which several indications of a warped disc have been found \citep{Marino2015,Casassus2015,Bohn2022}. 

For several discs with higher inclinations the vertical structure becomes visible in the isovelocity curves bending away from the semi-major axis in one direction (e.g. AA\,Tau, GM\,Aur, HD\,97048, LkCa\,15, PDS\,70, RY\,Lup, T\,Cha, V1366\,Ori). Fitting for this structure can be used to determine the flaring and scale height of the disc (e.g. \citealp{Casassus2019, Teague2019a}). Other more face-on or inclined but less elevated discs show a dipole morphology that is symmetric about the semi-major axis (e.g. AB\,Aur, CS\,Cha, HD\,135344B, HD\,139614, TW\,Hya, V4046\,Sgr).

To reveal possible deviations from Keplerian rotation that may be indicative of the presence of companions, we attempt to fit a Keplerian model to the rotation velocity of the discs in \hyperref[sec:velocityRes]{Sect.~\ref*{sec:velocityRes}} assuming thin and thick disc geometries.  
\begin{figure*}
\centering
\includegraphics[angle=90,width=0.95\textwidth]{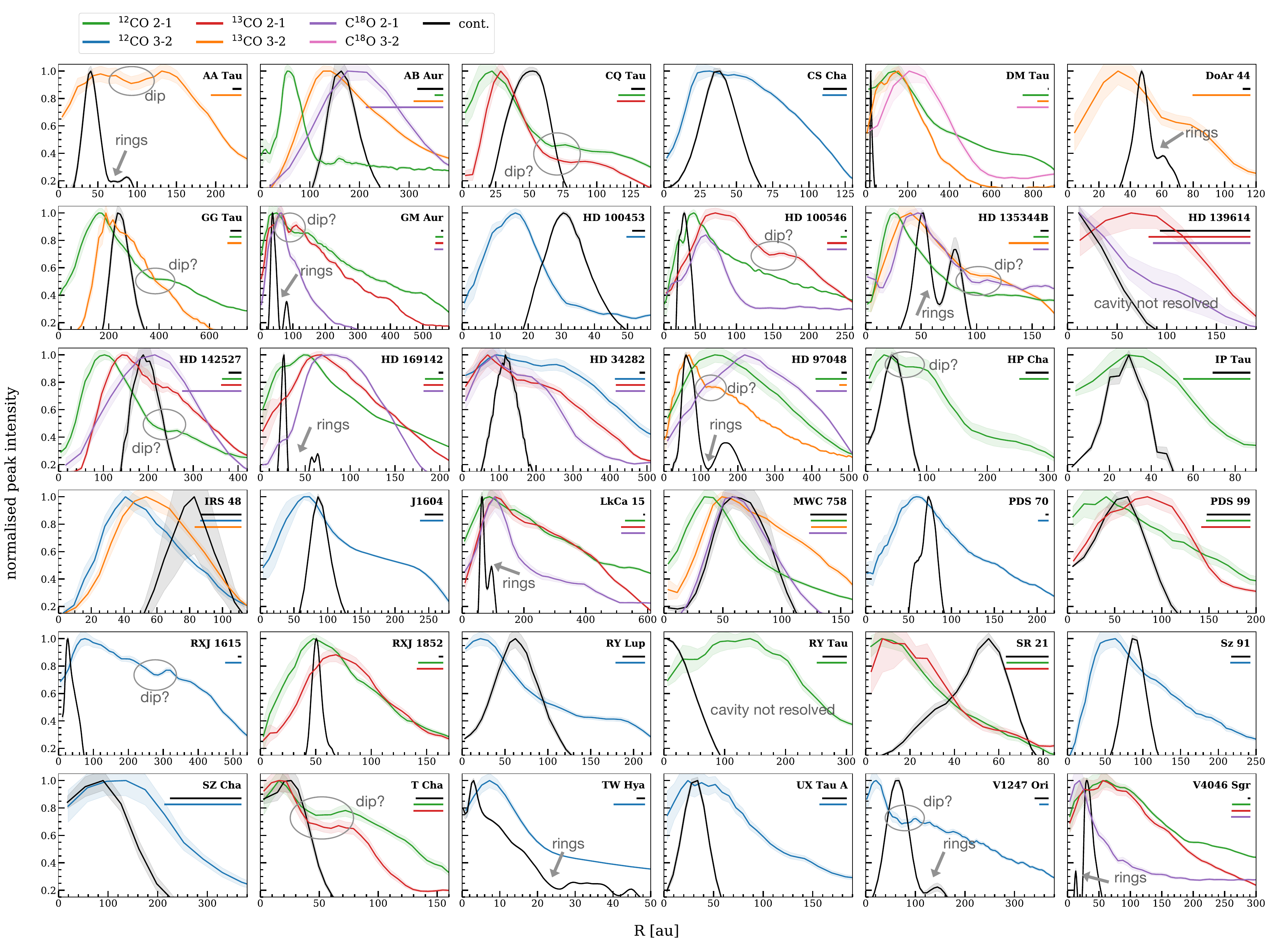}
\caption{Azimuthally averaged and normalised radial peak intensity profiles for the different CO lines (colored lines) and continuum (black lines) emission. Here blue corresponds to $^{12}$CO 3-2, green to $^{12}$CO 2-1, orange to $^{13}$CO 3-2, red to $^{13}$CO 2-1, pink to C$^{18}$O 3-2 and purple to C$^{18}$O 2-1. The major beam of each observation is indicated by the bars in the top right corner. Some features of the profiles are annotated in the individual panels.}\label{fig:radialProfiles}
\end{figure*}
\subsection{Radial profiles}\label{sec:radprof}
In \hyperref[fig:radialProfiles]{Fig.~\ref*{fig:radialProfiles}} the radial peak intensities are displayed for the different CO lines (colored lines) as well as the mm-continuum (black lines). These curves are calculated by azimuthally averaging the peak intensities for radial annuli of equal width, using the \texttt{GoFish} package \citep{Gofish}. By default, the widths of the annuli in this package are given as 1/4 of the beam major axis. For the computation we assume the geometries (thin or thick disc) obtained from the fitting of the rotation maps (see \hyperref[sec:velocityRes]{Sect.~\ref*{sec:velocityRes}}). For both geometries we recover similar radial profiles (due to similar fits, see discussion in \hyperref[sec:velocityRes]{Sect.~\ref*{sec:velocityRes}}) and thus the curves are only shown for the thin disc geometry in \hyperref[fig:radialProfiles]{Fig.~\ref*{fig:radialProfiles}}. All profiles are normalised to the peak value. The uncertainties, shown as shaded regions, correspond to the standard deviation per annulus divided by the square root of the number of independent beams in the annulus. Some discs in our sample are affected by cloud absorption, which can result in an artificially lower brightness temperature and a larger azimuthal scatter. We therefore exclude the affected azimuthal angles from our calculation. The beam size of the continuum and lines is indicated by a colored bar in each panel. The interpretation of radial profiles depends on the resolution. Given the inhomogeneity of our sample in that regard, the trends reported below may be subject to change with higher and comparable resolutions. The main features of the profiles are annotated in the individual panels of \hyperref[fig:radialProfiles]{Fig.~\ref*{fig:radialProfiles}}.       

The radial profiles can be used to estimate the size of the cavity (steep emission drop). In this context it is important to note that artificial cavities can be created in the peak intensity, depending on the beam size. Near the star the emission is less extended than the beam size and during the beam convolution the intensity gets diluted over the full extent of the beam, hence the peak intensity decreases. When studying the inner disc it is therefore important to look at the velocity integrated intensity, which is less affected by this issue. However, for the integrated intensity there may be contributions from the back side of the disc \citep{Rab2019}. The radial profiles for the integrated intensity are shown in the Appendix in \hyperref[fig:radialProfiles2]{Fig.~\ref*{fig:radialProfiles2}}. 

The measured brightness temperature will also be affected by the subtraction of continuum (except for HD\,97048 all data are continuum subtracted). For optically thick lines, that absorb part of the underlying continuum, line emission may be removed when subtracting the continuum, leading to artificial temperature drops (e.g. \citealp{Weaver2018, Rosotti2021, Bosman2021}). Since in this work we are mostly interested in substructures rather that obtaining a robust measurement of temperature, we do not expect this effect to significantly affect our results.

As visible in \hyperref[fig:radialProfiles]{Fig.~\ref*{fig:radialProfiles}} (and \hyperref[fig:radialProfiles2]{Fig.~\ref*{fig:radialProfiles2}}), the CO emission peaks inside the dust cavity for most discs as expected from previous work \citep{Bruderer2013,Marel2016}. For a few discs such as HD\,139614 the inner dust cavity is not resolved due to limited resolution. The radial profiles can further show dips or wiggles, especially in the dust indicating ring structures and depleted regions. It is important to note that it is also possible that instead of a dip, an enhanced desorption of CO ices by increased UV or a temperature inversion in the outer disc (more optically thin) at the edge of the continuum can result in an enhancement of gas emission (e.g. \citealp{Cleeves2016,Facchini2017}).
\begin{figure*}
\centering
\includegraphics[width=1.0\textwidth]{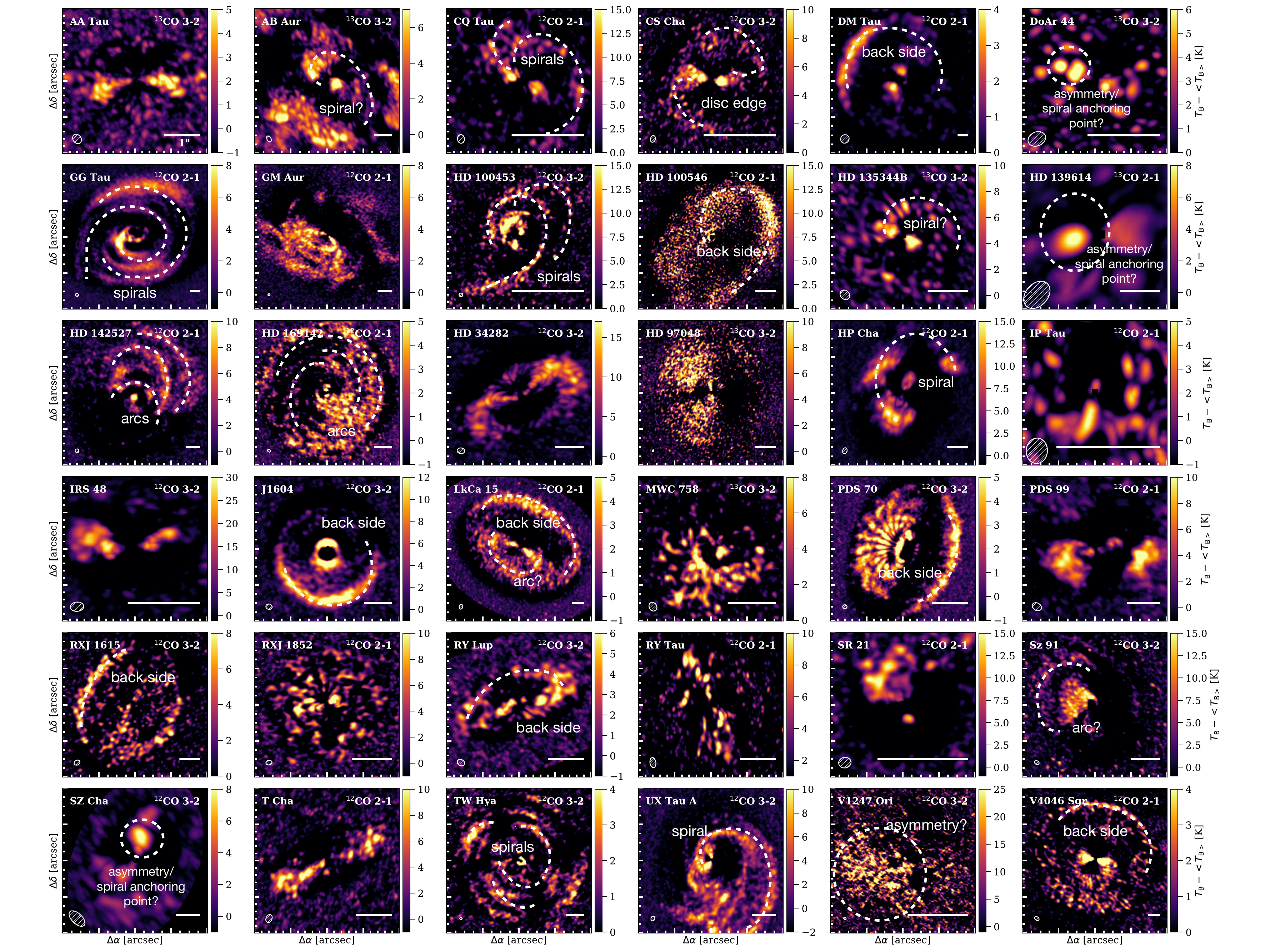}
\caption{Brightness temperature residuals shown for the main CO lines used in this analysis and for a geometrically thin disc approximation. Obtained with \texttt{GoFish}. The circle and bar in the bottom left and bottom right corner of each panel indicate the beam and a $1\arcsec$ scale respectively. The main features are annotated.}\label{fig:TbResMajorThin}
\end{figure*}
\section{Analysis}\label{sec:analysis}
\subsection{Brightness temperature structure}\label{sec:tempres}
To uncover possible substructures in the brightness temperature we construct an azimuthally symmetric model with \texttt{GoFish} and subtract this model from the data. The package computes an azimuthally averaged radial profile for a given geometry (compare \hyperref[sec:radprof]{Sect.~\ref*{sec:radprof}}) and then projects it onto the sky to create an azimuthally symmetric  model. For the disc geometry we use the results for the thin and (if available) the vertically extended thick disc from the kinematics modelling described in \hyperref[sec:velocityRes]{Sect.~\ref*{sec:velocityRes}}. The angles affected by cloud absorption are again excluded from the calculation. The resulting residuals are presented in \hyperref[fig:TbResMajorThin]{Fig.~\ref*{fig:TbResMajorThin}} for the geometrically thin disc and in the Appendix for the geometrically thick disc models (\hyperref[fig:thickboth]{Fig.~\ref*{fig:thickboth}}). The color scale is adapted such that regions with hotter temperatures than the model are highlighted. In this work we are mostly interested in the general occurrence of features such as spirals and therefore this choice was made to help the readability of the plot. Residuals for the other CO lines can also be found in the Appendix. The main features are annotated in the individual panels and further discussed in \hyperref[sec:Discussion]{Sect.~\ref*{sec:Discussion}}.   
\begin{figure*}
\centering
\includegraphics[width=1.0\textwidth]{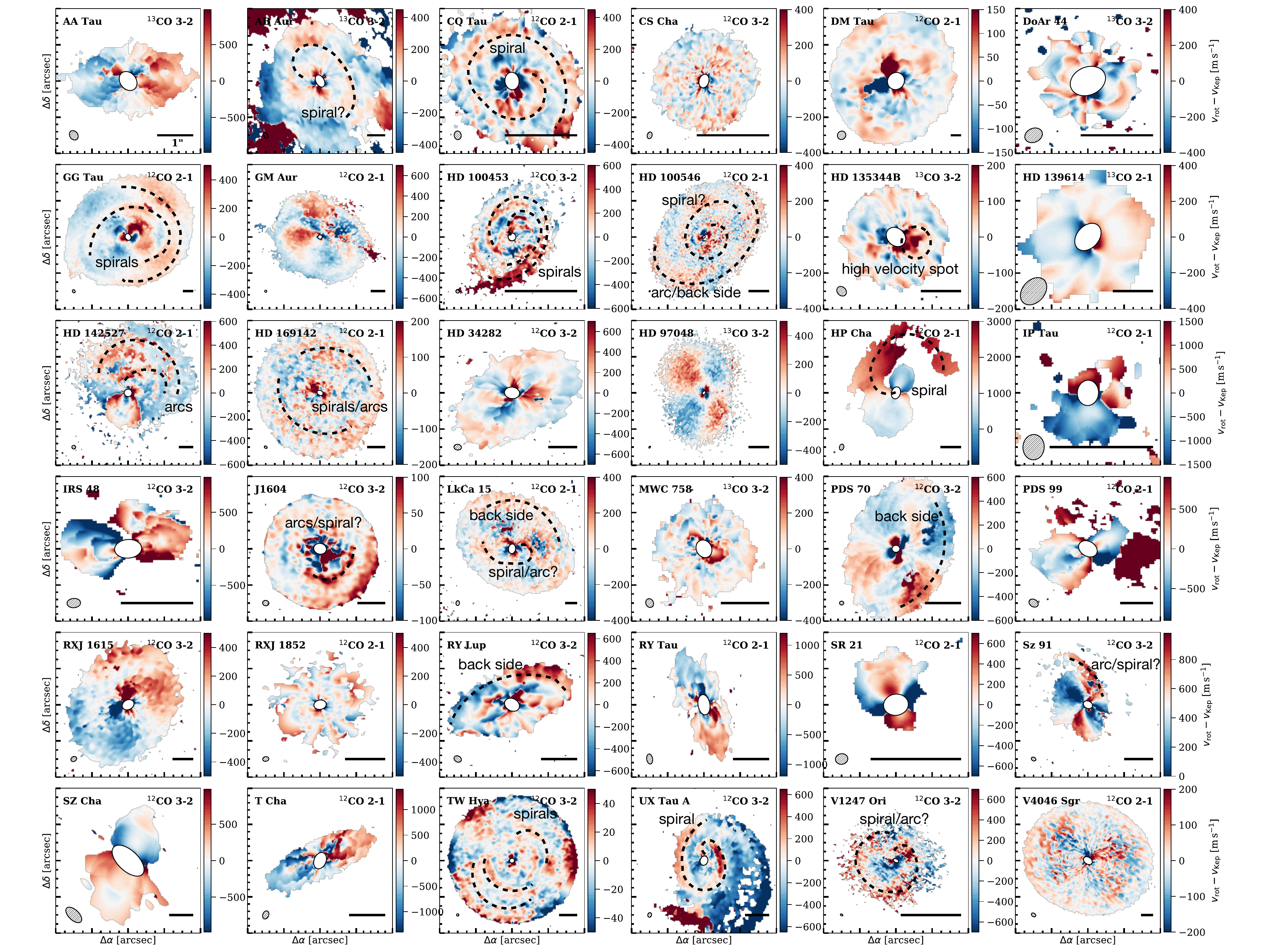}
\caption{Rotation velocity residuals shown for the main CO lines used in this analysis assuming a thin disc approximation. Obtained with \texttt{eddy}. The circle and bar in the bottom left and bottom right corner of each panel indicate the beam and a $1\arcsec$ scale respectively. The main features are annotated.}\label{fig:VrotResMajorThin}
\end{figure*}
\subsection{Velocity structure}\label{sec:velocityRes}
\begin{figure*}[h!]
\centering
\includegraphics[width=1.0\textwidth]{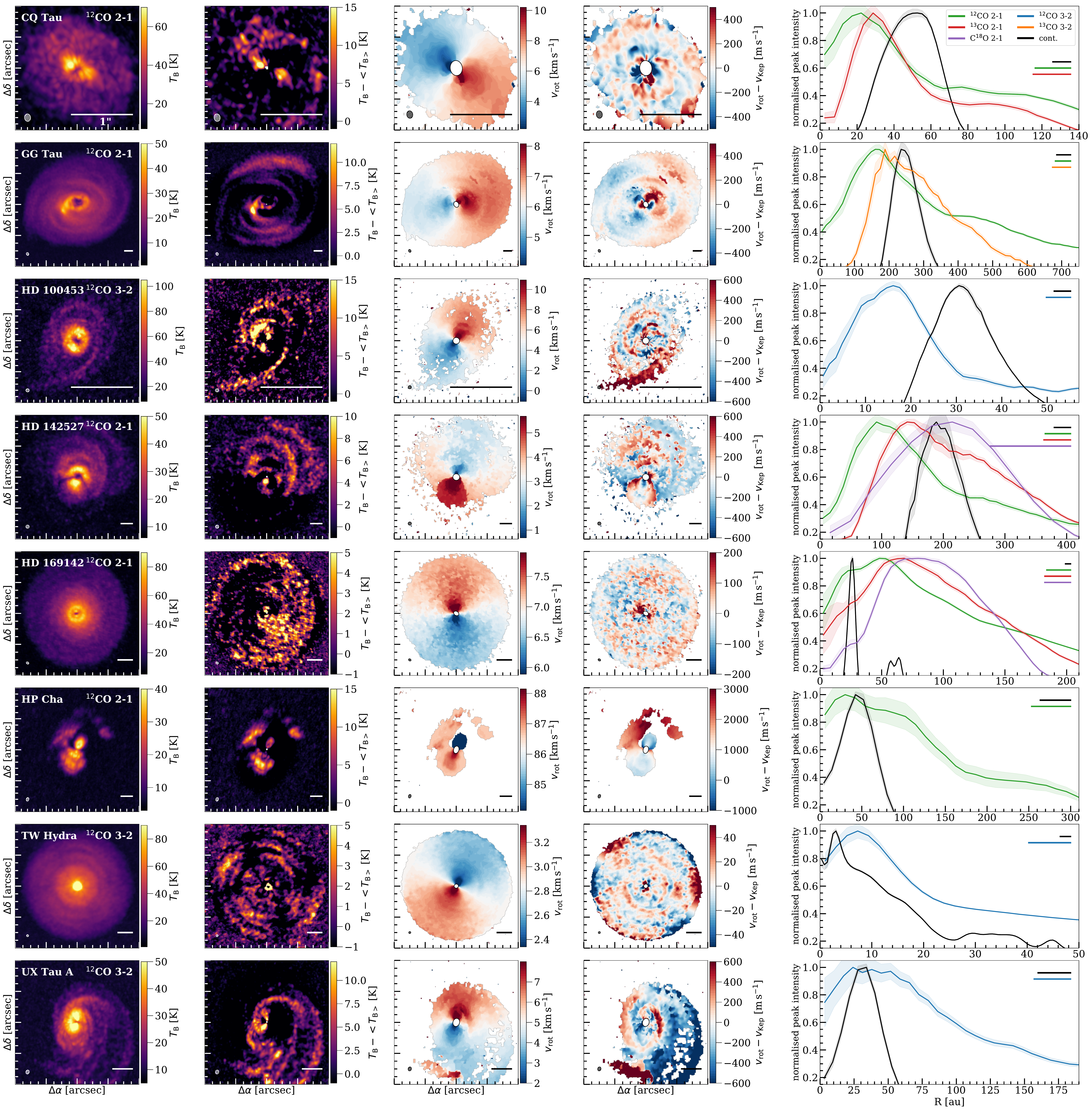}
\caption{Brightness temperature and velocity maps as well as their corresponding residuals and radial profiles, shown for the eighth targets with the clearest substructures. The circle and bar in the bottom left and bottom right corner of the maps indicate the beam and a $1\arcsec$ scale respectively. For the radial profiles, the major beam of each observations is indicated by the bars in the top right corner.}\label{fig:mapscomp}
\end{figure*}
To analyse the gas kinematics of our sample we use the \texttt{eddy} code \citep{eddy} to fit a Keplerian profile
\begin{equation}\label{eq:kepler}
v_{\mathrm{rot}} (r, \phi) = \sqrt{\frac{G M_*}{r}}  \cdot \cos{\phi} \cdot \sin{i} + v_{\mathrm{LSR}} 
,\end{equation}
with $(r,\phi)$ being the deprojected cylindical coordinates, $i$ the inclination of the disc and  $v_{\mathrm{LSR}}$ the systemic velocity, to the rotation maps shown in \hyperref[fig:vrotMapsMajor]{Fig.~\ref*{fig:vrotMapsMajor}}. To deproject the sky-plane coordinates $(x,y)$ into the midplane cylindrical coordinates $(r,\phi)$, the disc centre $(x_0,y_0)$, $i,$ and the disc position angle PA are used. The latter is measured between the north and the redshifted semi-major axis in an easterly direction. As a first step, the starting positions of the free fit parameters are optimised with \texttt{scipy.optimize} and their posterior distributions are estimated using the MCMC sampler. 

Besides the model for a geometrically thin disc, \texttt{eddy} also includes the possibility to fit for the vertical structure of the disc. To parameterise the emission layer, we choose a simple model of a flared disc described by
\begin{equation}\label{eq:scaleheight}
{z(r) = z_0 \cdot \left(\frac{r}{1\arcsec}\right) \cdot r^{\psi}}
\end{equation}
, where $z_0$ describes the elevation and $\psi$ the flaring angle of the emission surface.

In the modelling process we fix the object's distance and stellar mass, taken from the literature, and fit for the disc centre $(x_0, y_0 \in [-0\farcsec5,0\farcsec5])$, systemic velocity $(v_{\mathrm{LSR}} \in [v_{\mathrm{min}}\mathrm{(data)},v_{\mathrm{max}}\mathrm{(data)}])$, inclination $(i \in [-90\,\degree,90\,\degree]),$ and disc position angle (PA $\in [-360\,\degree,360\,\degree$]) as well as surface elevation $(z_0\in [0,5])$ and flaring angle $(\psi \in [0,5])$ in the geometrically thick disc approximation. We also conducted runs where the inclination was fixed instead of the stellar mass and where both the inclination and the stellar mass were left as free parameters. Overall the results are very similar for these different cases and in the following we will only show the results for the models where the stellar mass was fixed. 

For most targets we downsample the data by a factor of 2-4 before fitting. 
We fit for the whole disc, choosing an outer radius depending on the disc size to exclude possible noise at the disc edge and an inner radius of twice the beam major axis to exclude regions that are strongly affected by beam smearing. For a few cases, where the beam is very large, we slightly reduce this inner radius to ensure a reasonable number of pixels to fit. For all models we use 100 walkers, 8000 steps to burn in, and 2000 additional steps to sample the posterior distribution function and assume flat priors that are allowed to vary over a wide range. The uncertainties of the posterior distributions represent the 16th to 84th percentiles about the median value. The uncertainties on the kinematics, computed with \texttt{bettermoments} are included in the fit and shown in \hyperref[fig:uncer]{Fig.~\ref*{fig:uncer}}. They mostly lie well below the channel width but they increase in the central regions due to beam smearing. 

While most models converged rapidly within a few 100 steps, none of the models for HP\,Cha converged. Furthermore the models considering the vertical structure of the disc - even though rapidly converging - often do not match the bending of the isovelocity curves clearly seen in the data of the higher-inclination discs and return substantially smaller values for the elevation and flaring than expected. We tried both orientations of the inclination (positive and negative) in this context. For highly inclined sources the backside becomes prominently visible, resulting in a quadrupole morphology. However, this is fit by a dipole morphology and therefore the best fit lies between the two lobes of the quadrupole morphology, representing the average of the front and back side of the disc. The residuals of these models thus resemble those of the flat disc (see \hyperref[fig:VrotResMajorThin]{Fig.~\ref*{fig:VrotResMajorThin}} and \hyperref[fig:thickboth]{Fig.~\ref*{fig:thickboth}}). Higher spectral and spatial resolution data and individual modelling of the two disc sides (front and back) may be required to find a better fit for the vertical structure (e.g. directly from the channel maps). 

In \hyperref[fig:VrotResMajorThin]{Fig.~\ref*{fig:VrotResMajorThin}} the residuals after subtraction of the Keplerian model from the data are shown for the geometrically thin disc approximation. The residuals of the vertically extended disc approximation and the additional CO lines are presented in \hyperref[fig:thickboth]{Fig.~\ref*{fig:thickboth}}. Again, the residuals for the other CO lines can be found in the Appendix. In the plots we mask out disc regions inside of twice the beam, since these are strongly affected by beam smearing and not included in the fit. The main features are again annotated in the individual panels and further discussed in the following section. Same as for the brightness temperature residuals we mark the substructures found in the positive residuals. While it is interesting to study both the positive and negative residuals, the interpretation of these structures is not straight forward and a lot of effort is currently put into understanding the different patterns, which will be part of up-coming works. In contrast to that, this work aims at comparing the different substructures found in the gas of circumstellar discs that may be indicative of embedded planets.
\begin{table*}
\centering
\caption{Summary of the various features exhibited by our targets, that may be indicative of embedded planets\tablefootmark{a}.}
\begin{tabular}{l cccccccc}
\hline
\hline
source & deep gas & $T_{\mathrm{B}}$ & $v_{\mathrm{rot}}$ & NIR & NIR & Misalignment\tablefootmark{b} & comments & Ref.\tablefootmark{c} \\
& cavity & spirals/arcs & spirals/arcs & spirals & shadows & inner/outer\\
\hline
AA\,Tau & \color{Red}{\xmark} & \color{Red}{\xmark} & \color{Red}{\xmark} & - & - & - \\
AB\,Aur & \color{Green}{\cmark} & \color{Green}{(\cmark)} & \color{Green}{(\cmark)} & \color{Green}{\cmark} & \color{Red}{\xmark} & {\color{Green}{\cmark}} (A) & & 1\\
CQ\,Tau & \color{Green}{\cmark} & \color{Green}{\cmark} & \color{Green}{\cmark} & \color{Green}{\cmark} & \color{Green}{\cmark} & {\color{Green}{\cmark}} (A+G) & & 2 \\
CS\,Cha & \color{Green}{\cmark} & \color{Red}{\xmark} & \color{Red}{\xmark} & \color{Red}{\xmark} & \color{Red}{\xmark} & - & & 3\\
DM\,Tau & \color{Red}{\xmark} & \color{Red}{\xmark} & \color{Red}{\xmark} & - & - & - & low res.\\
DoAr\,44 & \color{Green}{\cmark} & \color{Green}{(\cmark)} & \color{Red}{\xmark} & \color{Red}{\xmark} & \color{Green}{\cmark} & - & & 2 \\
GG\,Tau & \color{Green}{\cmark} & \color{Green}{\cmark} & \color{Green}{\cmark} & \color{Green}{\cmark} & \color{Green}{\cmark} & {\color{Green}{\cmark}} (A) & multiple & 4\\
GM\,Aur & \color{Green}{\cmark} & \color{Red}{\xmark} & \color{Red}{\xmark} & \color{Red}{\xmark} & \color{Red}{\xmark} & {\color{Red}{\xmark}} (A+G) & & 2\\
HD\,100453 & \color{Green}{\cmark} & \color{Green}{\cmark} & \color{Green}{\cmark} & \color{Green}{\cmark} & \color{Green}{\cmark} & {\color{Green}{\cmark}} (A+G) & binary & 2\\
HD\,100546 & \color{Green}{\cmark} & \color{Red}{\xmark} & \color{Green}{(\cmark)} & \color{Green}{\cmark} & \color{Red}{\xmark} & - & & 2,5 \\
HD\,135344B & \color{Green}{\cmark} & \color{Green}{(\cmark)} & \color{Green}{(\cmark)} & \color{Green}{\cmark} & \color{Green}{\cmark} & - & & 2\\
HD\,139614 & \color{Red}{\xmark} & \color{Green}{(\cmark)} & \color{Red}{\xmark} & \color{Red}{\xmark} & \color{Green}{\cmark} & - & low res. & 2\\
HD\,142527 & \color{Green}{\cmark} & \color{Green}{\cmark} & \color{Green}{\cmark} & \color{Green}{\cmark} & \color{Green}{\cmark} & {\color{Green}{\cmark}} (A+G) & binary & 2\\
& & & & & & & warp, cloud absorption\\
HD\,169142 & \color{Green}{\cmark} & \color{Green}{\cmark} & \color{Green}{\cmark} & \color{Red}{\xmark} & \color{Green}{(\cmark)} & - & & 2,6 \\
HD\,34282 & \color{Green}{(\cmark)} & \color{Red}{\xmark} & \color{Red}{\xmark} & \color{Red}{\xmark} & \color{Red}{\xmark} & {\color{Green}{\cmark}} (A+G) & & 2\\
HD\,97048 & \color{Green}{\cmark} & \color{Red}{\xmark} & \color{Red}{\xmark} & \color{Red}{\xmark} & \color{Red}{\xmark} & - & cloud absorption & 2\\
HP\,Cha & \color{Red}{\xmark} & \color{Green}{\cmark} & \color{Green}{\cmark} & - & - & {\color{Green}{\cmark}} (A) & cloud absorption\\
IP\,Tau & \color{Red}{\xmark} & \color{Red}{\xmark} & \color{Red}{\xmark} & \color{Red}{\xmark} & \color{Red}{\xmark} & {\color{Red}{\xmark}} (A+G) & cloud absorption & 2\\
IRS\,48 & \color{Green}{\cmark} & \color{Red}{\xmark} & \color{Red}{\xmark} & - & - & - & cloud absorption\\
J1604 & \color{Green}{\cmark} & \color{Red}{\xmark} & \color{Green}{(\cmark)} & \color{Red}{\xmark} & \color{Green}{\cmark} & - & & 7\\
LkCa\,15 & \color{Green}{(\cmark)} & \color{Green}{(\cmark)} & \color{Green}{(\cmark)} & \color{Red}{\xmark} &  \color{Green}{(\cmark)} & - & & 2,8\\
MWC\,758 & \color{Green}{\cmark} & \color{Red}{\xmark} & \color{Red}{\xmark} & \color{Green}{\cmark} & \color{Red}{\xmark} & {\color{Green}{\cmark}} (A) & & 9\\
PDS\,70 & \color{Green}{\cmark} & \color{Red}{\xmark} & \color{Red}{\xmark} & \color{Red}{\xmark} & \color{Red}{\xmark} & {\color{Red}{\xmark}} (A+G) , {\color{Green}{\cmark}} (A) & imaged planets & 2\\
PDS\,99 & \color{Green}{\cmark} & \color{Red}{\xmark} & \color{Red}{\xmark} & - & - & - & cloud absorption \\
RXJ\,1615 & \color{Red}{\xmark} & \color{Red}{\xmark} & \color{Red}{\xmark} & \color{Red}{\xmark} & \color{Red}{\xmark} & {\color{Red}{\xmark}} (A+G) & & 2\\
RXJ\,1852 & \color{Green}{\cmark} & \color{Red}{\xmark} & \color{Red}{\xmark} & \color{Red}{\xmark} & \color{Red}{\xmark} & - & & 10\\
RY\,Lup & \color{Red}{\xmark} & \color{Red}{\xmark} & \color{Red}{\xmark} & \color{Green}{(\cmark)} & \color{Red}{\xmark} & {\color{Green}{\cmark}} (A+G) & & 2,11\\
RY\,Tau & \color{Red}{\xmark} & \color{Red}{\xmark} & \color{Red}{\xmark} & \color{Red}{\xmark} & \color{Red}{\xmark} & - & & 12\\
SR\,21 & \color{Red}{\xmark} & \color{Red}{\xmark} & \color{Red}{\xmark} & \color{Green}{\cmark} & \color{Red}{\xmark} & - & cloud absorption & 13\\
Sz\,91 & \color{Green}{\cmark} & \color{Green}{(\cmark)} & \color{Green}{(\cmark)} & \color{Red}{\xmark} & \color{Red}{\xmark} & - & cloud absorption & 14\\
SZ\,Cha & \color{Red}{\xmark} & \color{Green}{(\cmark)} & \color{Red}{\xmark} & \color{Red}{\xmark} & \color{Green}{\cmark} & {\color{Red}{\xmark}} (A+G) & low res., cloud absorption & 2\\
T\,Cha & \color{Red}{\xmark} & \color{Red}{\xmark} & \color{Red}{\xmark} & \color{Red}{\xmark} & \color{Red}{\xmark} & - & & 6\\
TW\,Hya & \color{Red}{\xmark} & \color{Green}{\cmark} & \color{Green}{\cmark} & \color{Red}{\xmark} & \color{Red}{\xmark} & {\color{Green}{\cmark}} (A) & & 16\\
UX\,Tau\,A & \color{Red}{\xmark} & \color{Green}{\cmark} & \color{Green}{\cmark} & \color{Green}{\cmark} & \color{Red}{\xmark} & - & binary & 2\\
V1247\,Ori & \color{Red}{\xmark} & \color{Green}{(\cmark)} & \color{Green}{(\cmark)} & \color{Green}{\cmark} & \color{Green}{\cmark} & {\color{Green}{\cmark}} (A+G) & & 2\\
V4046\,Sgr & \color{Red}{\xmark} & \color{Red}{\xmark} & \color{Red}{\xmark} & \color{Red}{\xmark} & \color{Red}{\xmark} & {\color{Green}{\cmark}} (A) & binary & 17\\
\hline
\end{tabular}
    \tablefoot{
        \tablefoottext{a}{Green checkmarks point out a detection, red crosses a non-detection and brackets indicate if a detection is tentative. The absence of substructures does not necessarily imply the absence of a companion/planet but may be a resolution/sensitivity effect.}
        \newline
        \tablefoottext{b}{A: Obtained from ALMA continuum data in \cite{Francis2020}, A+G: Obtained from ALMA CO and GRAVITY data in \cite{Bohn2022}}.
        \newline
        \tablefoottext{c}{References for spirals and shadows observed in the NIR scattered light. 1) \cite{Boccaletti2020}; 2) \cite{Bohn2022} (original reference can be found in this paper); 3) \cite{Ginski2018}; 4) \cite{Keppler2020}; 5) \cite{Garufi2016}; 6) \cite{Pohl2017}; 7) \cite{Pinilla2018}; 8) \cite{Thalmann2016}; 9) \cite{Benisty2015}; 10) \cite{Villenave2019}; 11) \cite{Langlois2018}; 12) \cite{Takami2013}; 13) \cite{MuroArena2020}; 14) \cite{Tsukagoshi2014}; 15) \cite{Pohl2017}; 16) \cite{Boer2020}; 17) \cite{Avenhaus2018}    
        }
        }
\label{tab:checks}
\end{table*}
\section{Discussion}\label{sec:Discussion}
The residuals presented in \hyperref[fig:TbResMajorThin]{Fig.~\ref*{fig:TbResMajorThin}} and \hyperref[fig:VrotResMajorThin]{Fig.~\ref*{fig:VrotResMajorThin}} show various features that are annotated in the different panels. For the discs that show very clear substructures (see \hyperref[sec:cleafFeat]{Sec.~\ref*{sec:cleafFeat}}), the different maps and radial profiles are collected again in \hyperref[fig:mapscomp]{Fig.~\ref*{fig:mapscomp}}.

For several higher-inclination discs, the vertical structure is still clearly visible in the residuals in form of a butterfly-like pattern even after subtraction of a geometrically thick disc model (e.g. AA\,Tau, HD\,34282). As mentioned before, most of these fits returned only slightly elevated emission surfaces and are not able to pick up the actual vertical structure of the disc. For these geometrically thick disc models still only a single side of the disc is modeled, while for the highly inclined discs both the front and back side are visible. An approach modelling both sides independently (such as used in the \texttt{Discminer}, \citealp{Izquierdo2021a}) would be needed to avoid this problem. Connected to that, in a few cases arc features can be seen at the disc edge, which result from looking into the back side of the disc (e.g J1604, LkCa\,15, PDS\,70). 

In \autoref{tab:checks} we list some features that may be indicative of interactions between disc and planets and/or stellar companions. Here a green checkmark stands for the detection of a feature, a red cross identifies a non-detection and brackets indicate a tentative detection. Especially if several features are observed in the same disc, this is a strong indication that we are tracing embedded planets/companions. As a deep gas cavity we mark cases where we see a clear drop in the radial profile of the peak and integrated intensity for at least the more optically thin lines that tend to trace the column density. In a few cases such as LkCa\,15 or HD\,34282 no steep drop is found but the gas cavity is very extended and better data is needed to confirm if a deep cavity is present \cite{Leemker2022}. In this work we do not investigate the presence of kink features and thus these are missing as a possible planet signpost in \autoref{tab:checks}. While \citep{Pinte2020} report the detection of such azimuthally located features in about half of the discs in the DSHARP program \citep{Andrews2018}, more data are needed to confirm the robustness of such claims and it is difficult to draw conclusions on the presence of kinks in our sample, given the inhomogeneity of spatial and spectral resolutions. Furthermore, the interpretation of kink features is not straight forward as they can be casued by a gap or density substructure rather than a planet \citep{Izquierdo2021a}. More detailed studies (beyond the scope of our work) of a homogenous data set at high resolution are needed to test for such scenarios. In the following we describe the observed structures in more detail.
\subsection{Clear spiral or arc like features}\label{sec:cleafFeat}
A few discs show clear spiral or arc like structures in both the brightness temperature and the velocity residuals (\hyperref[fig:mapscomp]{Fig.~\ref*{fig:mapscomp}}): CQ\,Tau \citep{Woelfer2021}, GG\,Tau, HD\,100453 \citep{Rosotti2020a}, HD\,142527 \citep{Garg2021}, HD\,169142, HP\,Cha, TW\,Hya \citep{Teague2019Spiral} and UX\,Tau\,A \citep{menard2020}. Half of these discs (CQ\,Tau, GG\,Tau, HD\,100453 and HD\,142527) are also marked by a spiral in the NIR, deep gas cavities, shadows in the NIR and a misalignment between inner and outer disc \citep{Francis2020, Bohn2022}. These four systems represent the best candidates for planet-disc or companion-disc interactions. In the cases of GG\,Tau and HD\,100453 and HD\,142527 binary components are indeed known, likely causing (at least part of) the observed spiral structures. Among these eight discs, HD\,142527 and HP\,Cha are affected by cloud absorption on the red-shifted and blue-shifted side respectively, however this is unlikely to explain the arc features on the blue-shifted side of HD\,142527 or the red-shifted spiral in HP\,Cha. 
\subsection{Tentative spiral or arc like features}
Some of the discs show tentative features (spirals, arcs or bright spots): For AB\,Aur, HD\,135344B (see also \citealp{Casassus2021}), LkCa\,15, Sz91 and V1247\,Ori these are found in both the brightness temperature and the velocity, with the ones in AB\,Aur and HD\,135344B - which are also marked by most other features - being most convincing. In addition to that, features are found in the brightness temperature residuals of DoAr\,44, HD\,139614 and SZ\,Cha and in the velocity residuals of HD\,100546 and J1604. Most of these discs have deep gas cavities and are marked by shadows in the NIR, for the other cases (e.g. HD\,139614 or SZ\,Cha) a deep cavity may be resolved with higher spatial resolution. Four of the ten sources with tentative features exhibit spirals in the NIR. It is important to note that some of the arc features may result from the misfit of the vertical structure (or other disc parameters) rather than dynamical interactions (see Fig. 12 in \citealp{Yen2020}). Furthermore some discs are strongly marked by cloud absorption. For example, the asymmetries seen in the brightness temperature of SZ\,Cha may result from this effect. For AB\,Aur, HD\,100546, HD\,139614, HD\,135344B and Sz\,91 the residuals of the additional lines (shown in the Appendix) partly support the detection of the described features.          
\subsection{No spiral/arc features}
Among the remaining sources that do not show any substructures in the kinematics or brightness temperature, most are in general marked by few of the indicators listed in \autoref{tab:checks} (except for a deep gas cavity). The only sources that have clearly observed NIR spirals are MWC\,758 and SR\,21. The lack of substructures in the brightness temperature and kinematics does not necessarily imply that there are no planet-disc interactions but these may be unresolved instead. Besides simply being an effect of missing spatial or spectral resolution, the position and mass of an embedded planet can also result in the absence of detectable substructures. For example, a planet closer to the star embedded in a deep cavity will be more difficult to trace with the chosen methods, that are mostly able to reveal substructures on larger scales. To confirm the presence or absence of features and identify different trends, follow-up, more homogeneous observations are needed. 
\subsection{Conclusive remarks}\label{sec:Conclusions}
Within our sample, eight discs are marked by clear substructures. In all cases features are seen in the gas temperature and the kinematics - indicating a connection - alongside other substructures in the gas and dust. Ten other discs show tentative features, of which half present signatures in the brightness temperature and kinematics simultaneously. Half of our sample is not showing any substructures in the ALMA data besides a deep gas cavity.

Except for MWC\,758 and SR\,21, all targets that exhibit clear spirals in NIR scattered light show at least tentative features in the brightness temperature and /or kinematics. Scattered light is tracing the hot upper disc layers, most likely to show spirals, and since $^{12}$CO is also tracing the disc surface we expect to pick up these features there as well \citep{Law2021}. However, many of the tentative features in our sample are not spirals, which is likely related to poor spatial resolutions as well as cloud absorption in some cases. To resolve the spirals expected from the NIR, high sensitivity and spectral and spatial resolution ALMA data is crucial. For the targets showing cloud absorption, deep observations of the less affected  $^{13}$CO may help to pick up clear substructures. Two targets (HP\,Cha and TW\,Hya) show clear spiral structures without a counterpart having yet been observed in the NIR. 

It is expected that substructures such as spirals are more likely found in systems with high luminosity and a wide cavity, where the upper disc layers can reach higher temperatures, making it much easier to observe them (e.g. \citealp{Garufi2018,vanderMarel2021}). We search for such correlations in our data set: While we do not find any clear trends, most discs that show no spiral substructures indeed represent the less massive, cooler and less luminous stars and features tend to become more visible for the more luminous sources. Given however the inhomogeneity of our sample we expect a large observational bias and to draw clear conclusions on correlations it is crucial to study a more uniform (in terms of resolution) data set, that includes a wide range of spectral types and stellar masses. As shown for TW\,Hya, for which very high sensitivity data exist at high spatial resolution ($\sim$ 8\,au), substructures can still be distinguished despite a sub-solar stellar mass and an almost face on disc. To understand if there exist differences between the different stellar groups comparable observations are essential. 

Connected to that, the lack of features can not be seen as an indication of missing embedded planets. PDS\,70 for example hosts two confirmed massive planets, that have been directly imaged, but does not show any other features in our data despite a deep gas cavity. This is likely because the planets are located further inside the cavity in this system and higher sensitivity plus resolution observations are needed to reveal substructures in the temperature and the kinematics, related to dynamical interactions.

Furthermore, as shown by \cite{Izquierdo2021a} an embedded planet has to be rather massive to excite strong observable signatures in the kinematics: From simulations the authors find strong perturbations for planets more massive than 1\,$M_{\mathrm{J}}$, however to pick these up in ALMA data, very high spectral and spatial resolution is essential. Thus it is not surprising that in our sample the clearest features are seen in the multiple star systems.    
%
%
%
\section{Summary}\label{sec:Summary}
In this work we have analysed the brightness temperatures and kinematics of a sample of 36 large cavity transition discs, representing the best candidates to search for dynamical interactions. Our main results are summarised as follows: 
\begin{itemize}
\item Eight discs out of our sample show significant perturbations in both the brightness temperature and velocity residuals, while no features are found in half of the sample at the current (spatial and spectral) resolution and sensitivity.
\vspace{0.1cm}
\item Several discs show tentative features that need to be confirmed with deep, high resolution ALMA observations in the upcoming years.
\vspace{0.1cm}
\item Almost all targets that exhibit spirals in NIR scattered light show at least tentative features in the CO data.
\vspace{0.1cm}
\item In most cases our method reveals deviations that are caused by sub-stellar companions.
\vspace{0.1cm}
\item 
For about 60\,\% of the sources a deep gas cavity is resolved in addition to the dust cavity at the current spatial resolution. 
\end{itemize}
To detect planets in the Jupiter-mass range the available observations are neither deep enough nor do they have the required spatial and spectral resolution, explaining the lack of features in many discs. Up-coming and future deep ALMA observations at high spectral and spatial resolution together with dedicated modeling efforts may reveal more of such features and help to disentangle different formation scenarios. 
\section*{Acknowledgements}
We greatly thank the referee Ruobing Dong for his helpful feedback that improved the quality of this work. We also would like thank all the people that kindly provided the reimaged/self-calibrated data sets. This paper makes use of different ALMA data sets, detailed in \autoref{tab:dataProp} and \autoref{tab:dataPropAdd}. ALMA is a partnership of ESO (representing its member states), NSF (USA) and NINS (Japan), together with NRC (Canada) and NSC and ASIAA (Taiwan) and KASI (Republic of Korea), in cooperation with the Republic of Chile. The Joint ALMA Observatory is operated by ESO, auI/NRAO and NAOJ. 
%
\bibliographystyle{aa}
\bibliography{bibliography}
\begin{appendix} 
\onecolumn
\section{Continuum}\label{appendix:Continuum}
\begin{figure*}[h!]
\centering
\includegraphics[width=1.0\textwidth]{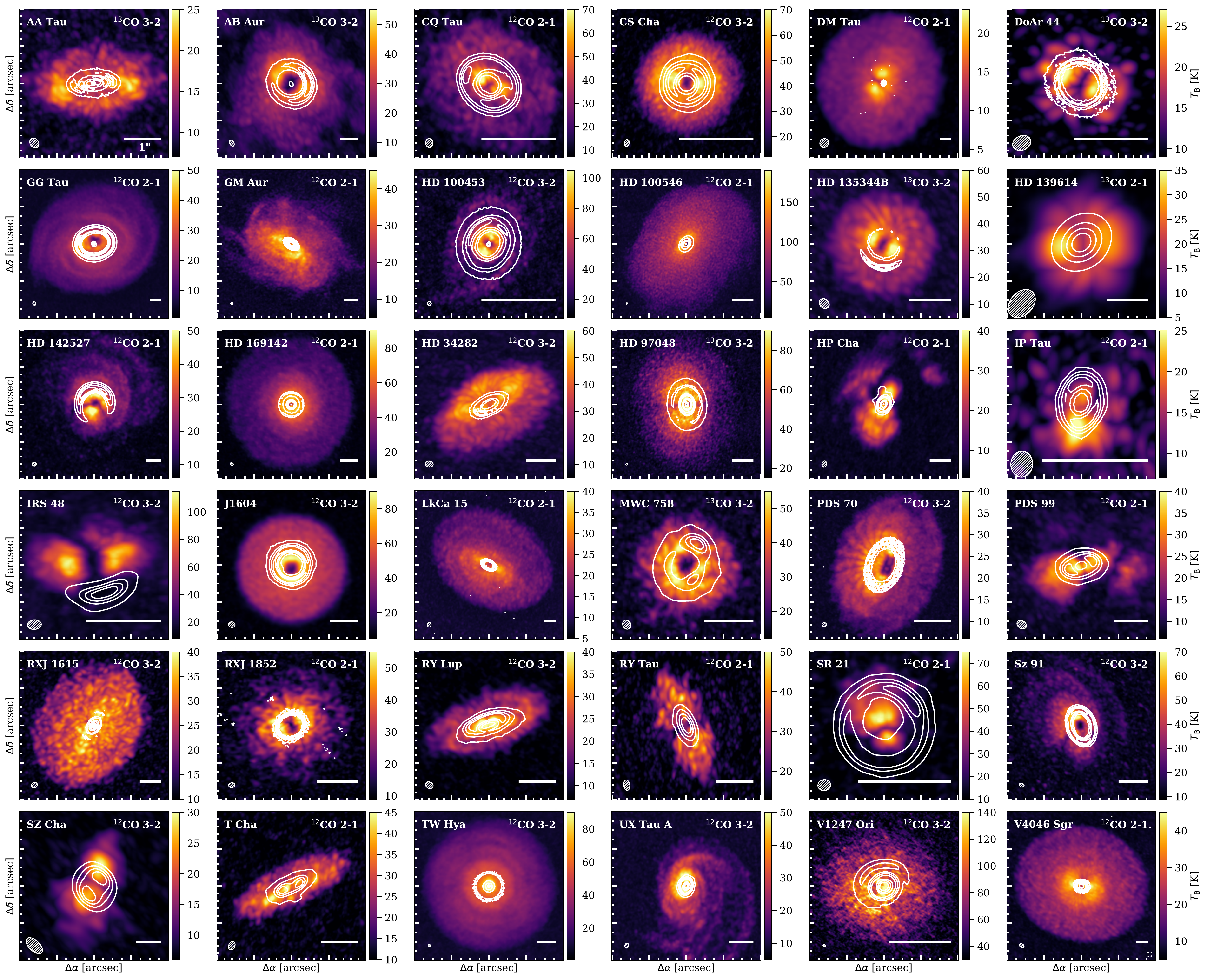}
\caption{Peak brightness temperature maps of the gas emission in our targets, shown for the main CO lines with the continuum overlaid as white contours. Five contour levels are equally spaced between $3\,\sigma$ and the peak flux of the continuum. The circle and bar in the bottom left and bottom right corner of each panel indicate the beam and a $1\arcsec$ scale respectively.}\label{fig:TbMapsMajorCont}
\end{figure*}

\hyperref[fig:TbMapsMajorCont]{Figure~\ref*{fig:TbMapsMajorCont}} shows the peak brightness temperature maps of the main lines used for analysis in this work with overlaid mm-continuum (either ALMA B6 or B7). For most cases, the size of the dust disc is substantially smaller than that of the gas disc.    
\newpage
\section{Thick disc residuals of the main lines}\label{appendix:thick}
\begin{figure*}[h!]
\centering
\includegraphics[width=1.0\textwidth]{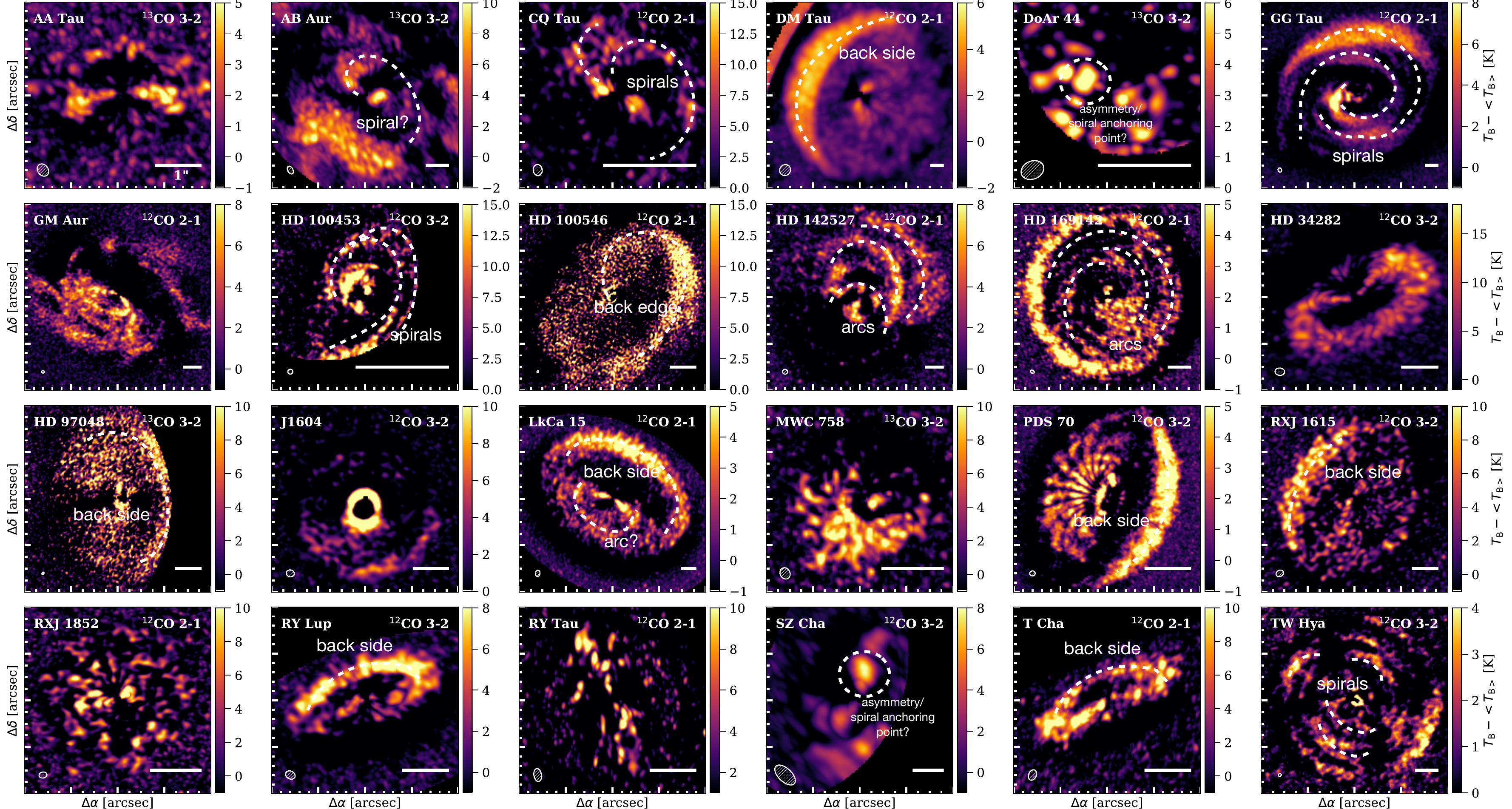}
\includegraphics[width=1.0\textwidth]{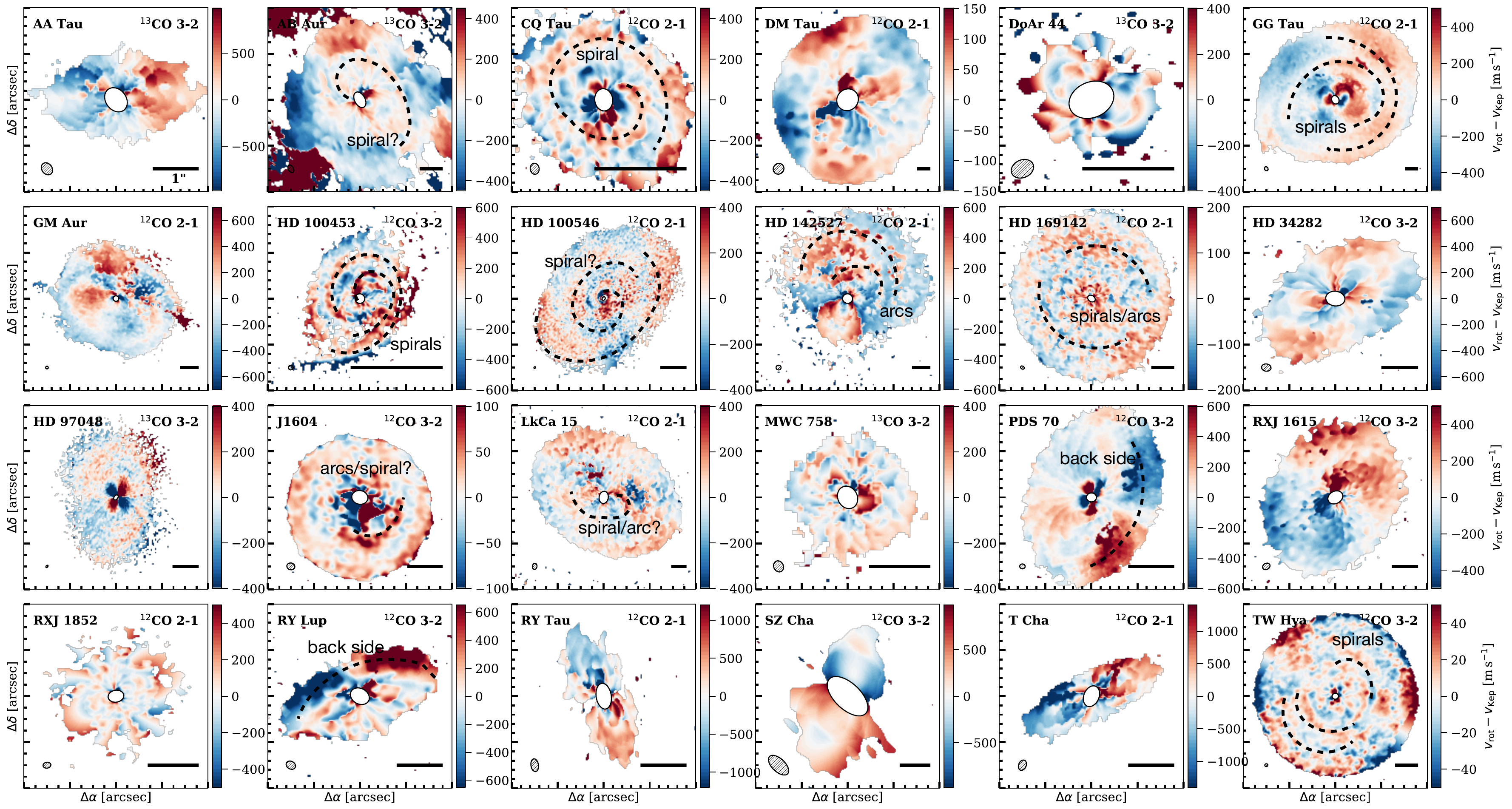}
\caption{Brightness temperature (top) and rotation velocity (bottom) residuals, shown for the main lines used in this analysis and for a thick disc geometry. The circle and bar in the bottom left and bottom right corner of each panel indicate the beam and a $1\arcsec$ scale respectively. Some features are annotated.}\label{fig:thickboth}
\end{figure*}
\newpage
\section{Uncertainties of the rotation velocity}
\begin{figure*}[h!]
\centering
\includegraphics[width=1.0\textwidth]{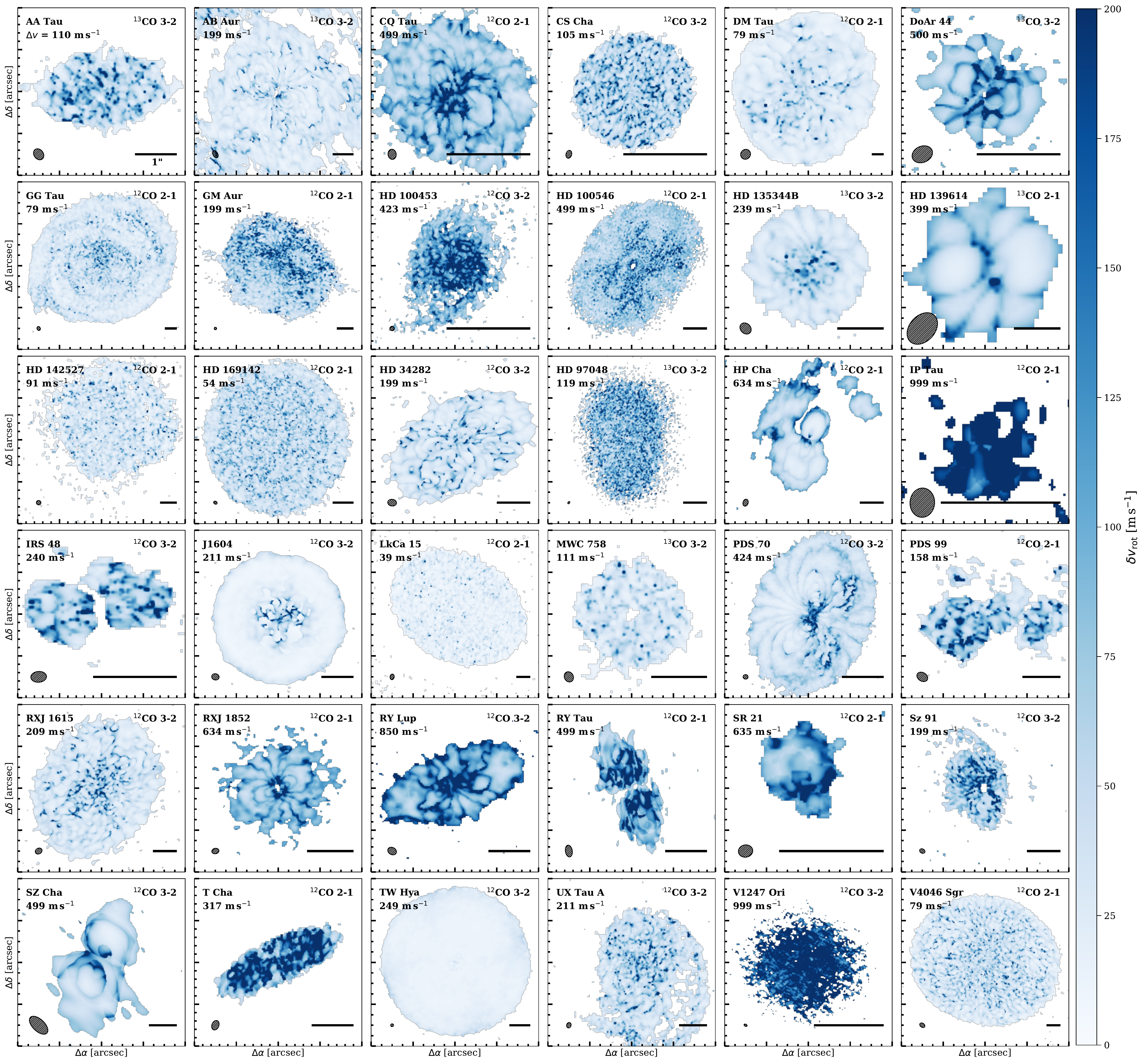}
\caption{Kinematical uncertainties of the main lines of this analysis. Computed with \texttt{bettermoments}. The circle and bar in the bottom left and bottom right corner of each panel indicate the beam and a $1\arcsec$ scale respectively. The channel spacing is shown in the top left corner of each panel.}\label{fig:uncer}
\end{figure*}

In \hyperref[fig:uncer]{Fig.~\ref*{fig:uncer}}
we show the uncertainties of the kinematics for the main lines, computed with \texttt{bettermoments}. These statistical uncertainties are calculated by linearising and propagating the uncertainty from the fluxes to the centroid estimate. They uncertainties mostly lie well below the channel width but increase in the central regions due to beam smearing. For lower sensitivity observations, thermal broadening plays an important role, significantly increasing the uncertainties.  
\newpage
\section{Comparison of reimaged and archival data products}\label{appendix:CompareCleaning}
\begin{figure*}[h!]
\centering
\includegraphics[width=0.7\textwidth]{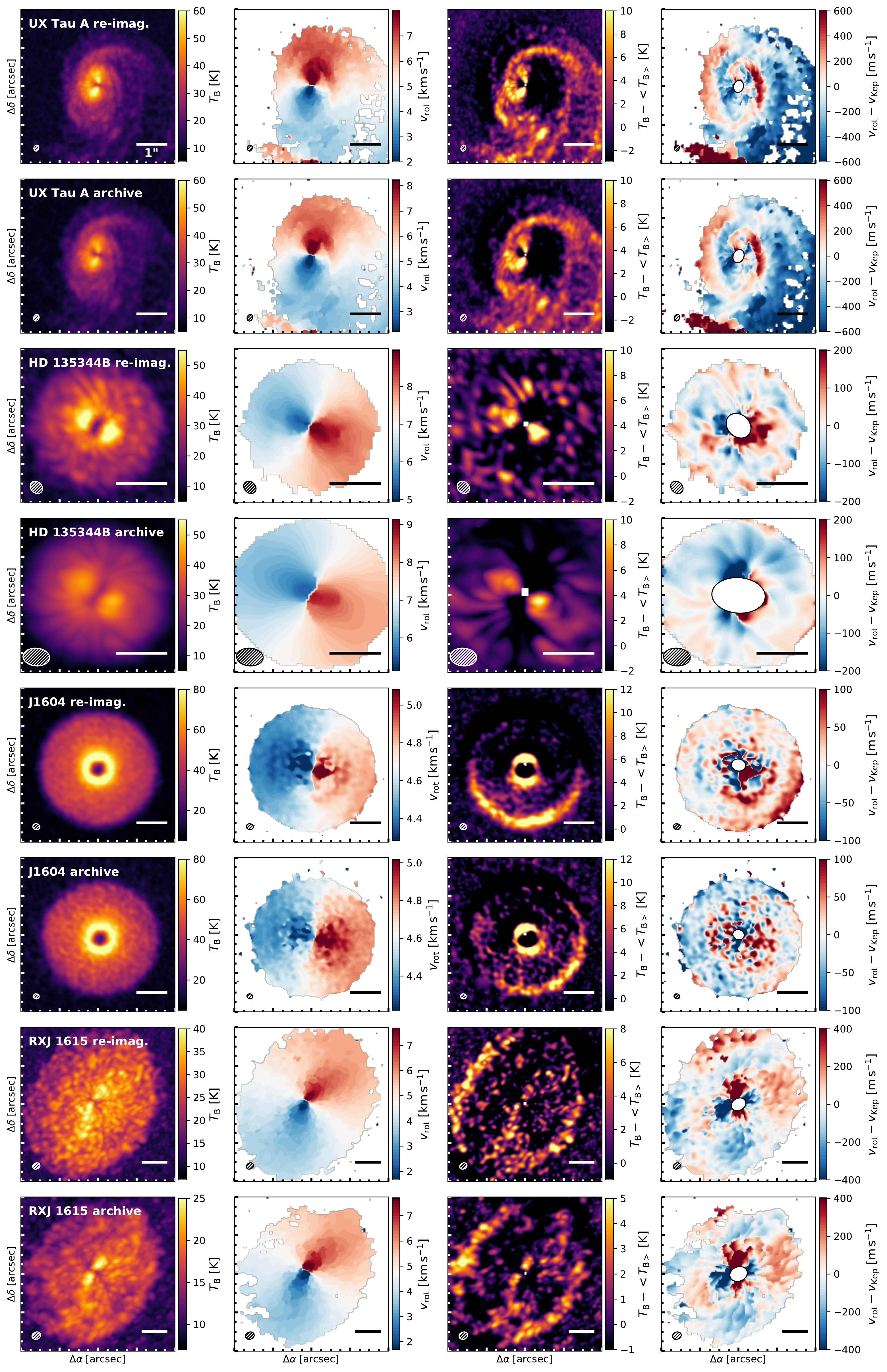}
\caption{Comparison between the reimaged and archival data for four sample sources, including one with clear spirals, two with tentative and one without any features. The circle and bar in the bottom left and bottom right corner of each panel indicate the beam and a $1\arcsec$ scale respectively.}\label{fig:compClean}
\end{figure*}

In this work we combine reimaged with archival data products and it is important to understand what impact this has on the results. For the data where reimaged sets are available, we conduct our analysis for both the reimaged and the according archival image cubes and compare the results. A few examples are shown in \hyperref[fig:compClean]{Fig.~\ref*{fig:compClean}} for a disc with clear spirals, two discs with tentative spirals and one without any observed features. We find that the fitting procedure is not significantly affected by the choice of cube. Moreover, the detection of clear spirals as well as a non-detection do not depend on the data set. Due to an increased S/R, tentative features come out stronger (e.g. J1604) and in some cases only become visible (e.g. HD\,135344B) in the reimaged products, thus some tentative substructures may be hidden in the cases that we classified as non-detections, when the archival product data was used.
\newpage
\section{Additional Lines}\label{appendix:AppendixAddLines}
This appendix comprises the results for the additional lines used in this work. When studying substructures in discs it is useful to look at various molecular tracers that probe different disc layers. Understanding if and how substructures vary vertically and radially can be used to assess the underlying pertrubation. For example, in a passively heated disc with a vertical temperature gradient, the opening angle of a spiral is expected to decrease towards the midplane, while for spirals launched by gravitational instability the midplane would be heated by shocks, resulting in similarly wound spirals throughout the disc \citep{Juhasz2018}. 
\subsection{Characteristics}\label{appendix:table}
Some characteristics of the data are listed below in \autoref{tab:dataPropAdd}. In most cases, the archival data products were used for analysis, as indicated in the last column.  
\begin{table*}[h!]
    \centering
    \caption{Characteristics of the additional ALMA line data}
    \begin{tabular}{llccccccc}
    \hline
    \hline
        Object & Line & ALMA Project ID & Beam & $\Delta \upsilon$ & RMS & Cube \tablefootmark{a} \\
        & & & ($\arcsec$) & (km$\,$s$^{-1}$) & (mJy$\,$beam$^{-1}$) & source \\
        \hline
        ABAur & $^{12}$CO 3-2 & 2012.1.00303.S & 0.31x0.19 & 0.05 & 11.4 & P/PC \\
        & $^{12}$CO 2-1 & 2015.1.00889.S & 0.11x0.08 & 0.32 & 3.5 & A\\ 
        & $^{13}$CO, C$^{18}$O 2-1 & 2019.1.00579.S & 0.95x0.57  & 0.17 & 9.5, 7.9 & A\\
        CQTau & $^{13}$CO 2-1 & 2013.1.00498.S & 0.13x0.1 & 0.7 & 0.9 & P/PC\\
        & & 2016.A.00026.S & & \\
        & & 2017.1.01404.S & & \\
        DMTau & $^{12}$CO 3-2 & 2013.1.00647.S & 1.0x0.75 & 0.2 & 55.9 & A\\ 
        & $^{13}$CO, C$^{18}$O 3-2 & 2016.1.00565.S & 0.37x0.29, 1.04x0.81 & 0.11 & 16.8, 21.4 & A\\ 
        & $^{13}$CO, C$^{18}$O 2-1 & 2016.1.00724.S & 0.9x0.84 & 0.08 & 19.2 , 14.1 & A\\
        GGTau & $^{12}$CO, $^{13}$CO 3-2 & 2012.1.00129.S & 0.4x0.3 & 0.25 & 5.7, 7.8 & A\\
        GMAur & $^{13}$CO 3-2 & 2016.1.00565.S & 0.38x0.26 & 0.11 & 16.2 & A\\
        & $^{13}$CO, C$^{18}$O 2-1 & 2018.1.01055.L & 0.15x0.15, 0.17x0.13 & 0.2 & 2.7, 1.1  & P/PC \\
        HD100546 & $^{12}$CO 3-2 & 2011.0.00863.S & 0.94x0.42 & 0.11 & 19.4 & A\\ 
        & $^{13}$CO, C$^{18}$O 2-1 & 2016.1.00344.S & 0.25x0.14 & 0.17 & 6.8, 6.4 & A\\
        HD135344B & $^{12}$CO 3-2 & 2012.1.00870.S & 0.36x0.29 & 0.11 & 38.9 & A\\
        & $^{12}$CO, $^{13}$CO, C$^{18}$O 2-1 & 2018.1.01066.S & 0.1x0.08 & 0.2 & 2.6, 2.8, 2.4 & P/PC \\
        HD139614 & C$^{18}$O 2-1 & 2015.1.01600.S & 0.73x0.53 & 0.33 & 11.9 & A\\
        HD142527 & $^{12}$CO 3-2 & 2011.0.00465.S & 0.57x0.35 & 0.5 & 11.1 & P/PC\\ 
        & $^{13}$CO, C$^{18}$O 3-2 & 2012.1.00725.S & 0.31x0.27 & 0.11 & 8.0, 9.9 & A\\
        & $^{13}$CO, C$^{18}$O 2-1 & 2015.1.01353.S & 0.29x0.26, 0.84x0.77 & 0.1 & 7.5, 19.3 & A\\
        HD169142 & $^{12}$CO, $^{13}$CO 3-2 & 2012.1.00799.S & 0.18x0.13, 0.19x0.13 & 0.21, 0.22 & 17.9, 19.8  & A\\ 
        & $^{13}$CO, C$^{18}$O 2-1 & 2015.1.00490.S & 0.19x0.14 & 0.08 & 5.4, 4.0 & A\\ 
        HD34282 & $^{12}$CO, $^{13}$CO, C$^{18}$CO 2-1 & 2015.1.00192.S & 0.24x0.21, 0.25x0.23 & 0.08, 0.17 & 10.6, 7.9, 5.5 & A\\
        HD97048 & $^{12}$CO 3-2 & 2013.1.00658.S & 0.65x0.39 & 0.2 & 13.93 & A\\
        & $^{12}$CO, $^{13}$CO, C$^{18}$O 2-1 & 2015.1.00192.S & 0.46x0.22 & 0.3 & 6.9, 6.5, 6.30, 4.9 & P/PC\\
        HPCha & $^{12}$CO 3-2 & 2013.1.01075.S & 0.77x0.4 & 0.2 & 35.7 & A\\ 
        IRS48 & $^{13}$CO 3-2 & 2013.1.00100.S & 0.21x0.16 & 0.26 & 14.4 & A\\ 
        LkCa15 & $^{13}$CO, C$^{18}$O 2-1 & 2018.1.00945.S & 0.48x0.3 & 0.33 & 3.7, 4.8 & P/PC\\ 
        MWC758 & $^{12}$CO 3-2 & 2011.0.00320.S & 0.82x0.47 & 0.05 & 37.3 & A\\ 
        & C$^{18}$O 3-2 & 2012.1.00725.S & 0.34x0.22 & 0.6 & 26.1 & A\\
        & $^{12}$CO, $^{13}$CO, C$^{18}$O 2-1 & 2017.1.00940.S &  0.2x0.15 & 1.26, 1.33 & 1.7, 1.6, 1.2& A\\ 
        PDS99 & $^{13}$CO 2-1 & 2015.1.01301.S & 0.34x0.23 & 0.17 & 9.8 & A\\ 
        RXJ1852 & $^{13}$CO 2-1 & 2018.1.00689.S & 0.16x0.12 & 0.66 & 4.8 & A\\ 
        RYLup & $^{12}$CO 2-1 & 2017.1.00449.S & 0.19x0.16 & 0.04 & 15.7 & A\\
        SR21 & $^{13}$CO 3-2 & 2012.1.00158.S & 0.27x0.23 & 0.24 & 11.5 & A\\
        & $^{13}$CO 2-1 & 2018.1.00689.S & 0.15x0.12 & 0.66 & 5.1 & A\\
        Sz91 & $^{12}$CO 2-1 & 2013.1.00663.S & 0.62x0.59 & 0.5 & 27.6 & A\\
        TCha & $^{12}$CO, $^{13}$CO 3-2 & 2012.1.00182.S & 0.26x0.14, 0.27x0.14 & 0.85, 0.88 & 14.5, 17.4  & A\\
        & $^{13}$CO 2-1 & 2017.1.01419.S & 0.24x0.16 & 1.33 & 5.0 & A\\
        UXTauA & $^{12}$CO 2-1 & 2013.1.00498.S & 0.26x0.21 & 0.63 & 7.4 & A\\
        V4046Sgr & $^{12}$CO, $^{13}$CO 3-2 & 2016.1.00315.S & 0.28x0.17, 0.3x0.18 & 0.21, 0.44 & 25.4, 22.5 & A\\ 
        & $^{13}$CO, C$^{18}$CO 2-1 & 2016.1.00724.S & 0.42x0.3 & 0.08 & 11.4, 7.8  & A\\
        \hline
    \end{tabular}
        \tablefoot{
        \tablefoottext{a}{P/PC: Reimaged data cube. Public data or obtained via private communication, A: Archival data product.}
        }
    \label{tab:dataPropAdd}
\end{table*}
\newpage
\subsection{Brightness temperature maps}\label{appendix:tbmaps}
\begin{figure*}[h!]
\centering
\includegraphics[width=1.0\textwidth]{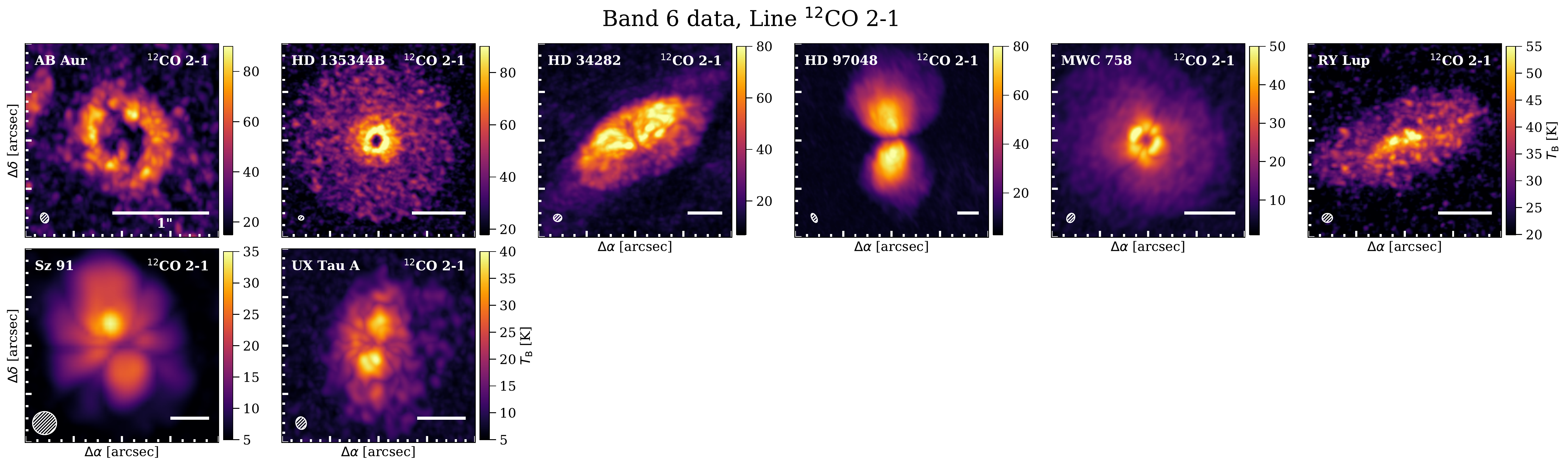}
\includegraphics[width=1.0\textwidth]{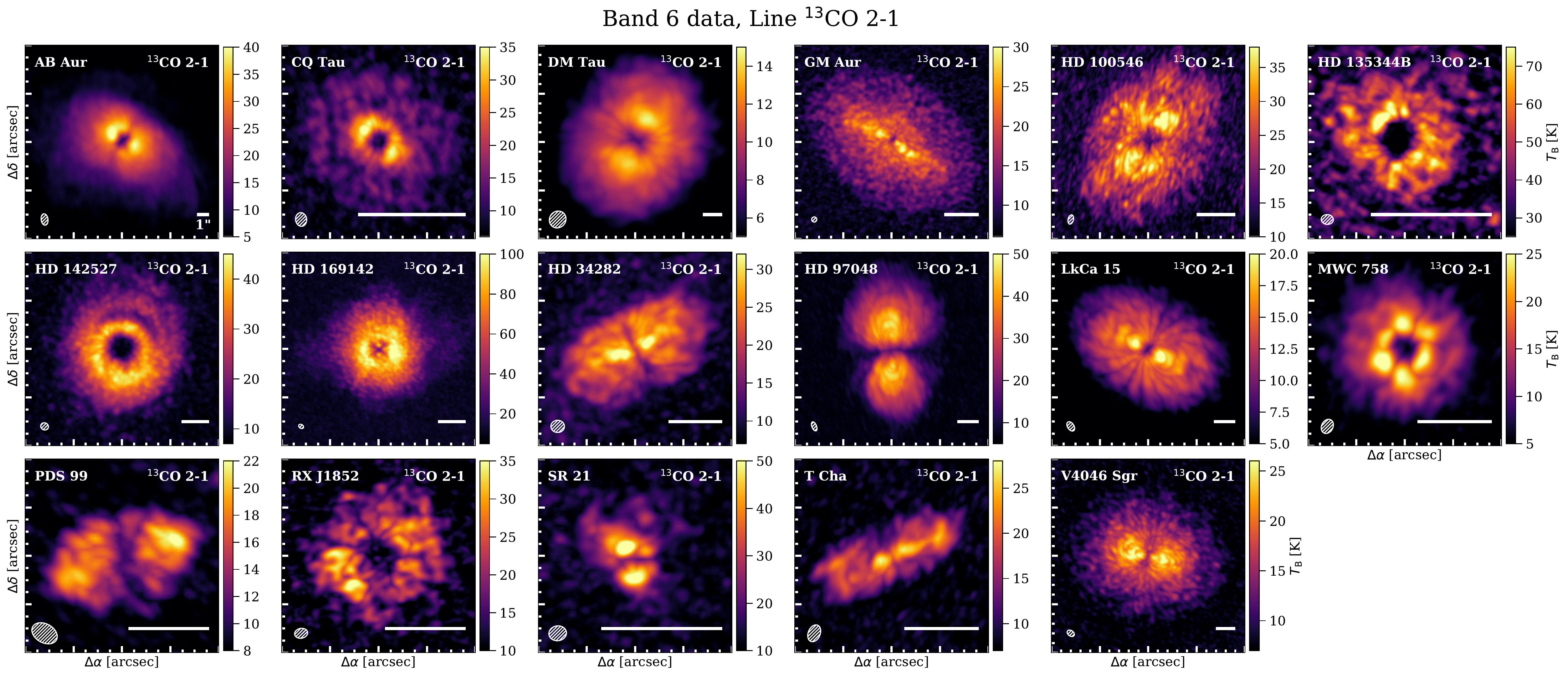}
\includegraphics[width=1.0\textwidth]{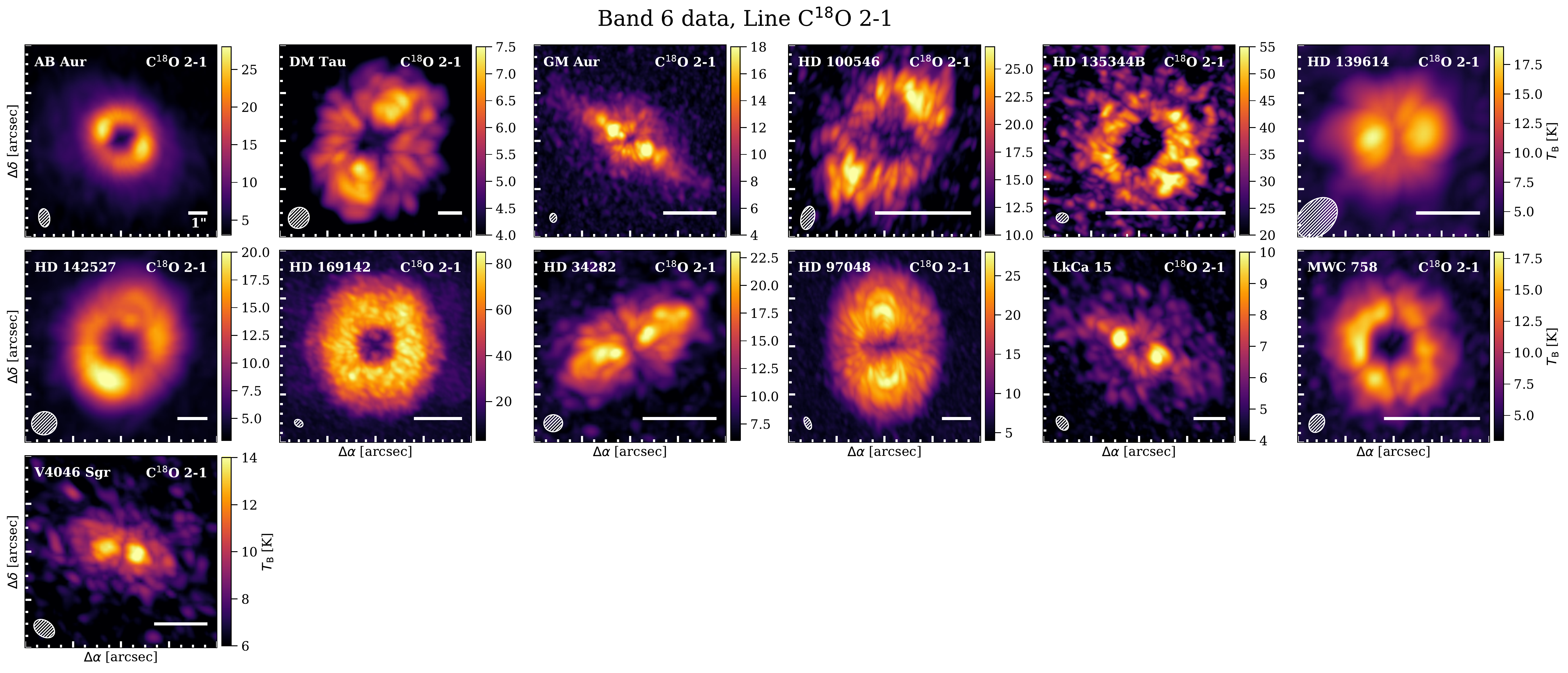}
\caption{Peak brightness temperature maps of the gas emission in our targets, shown for the additional Band 6 CO lines used in this analysis. The conversion from peak intensity to units of Kelvin is done with the Planck law. The circle and bar in the bottom left and bottom right corner of each panel indicate the beam and a $1\arcsec$ scale respectively.}\label{fig:TbAddBand6}
\end{figure*}

\begin{figure*}[h!]
\centering
\includegraphics[width=1.0\textwidth]{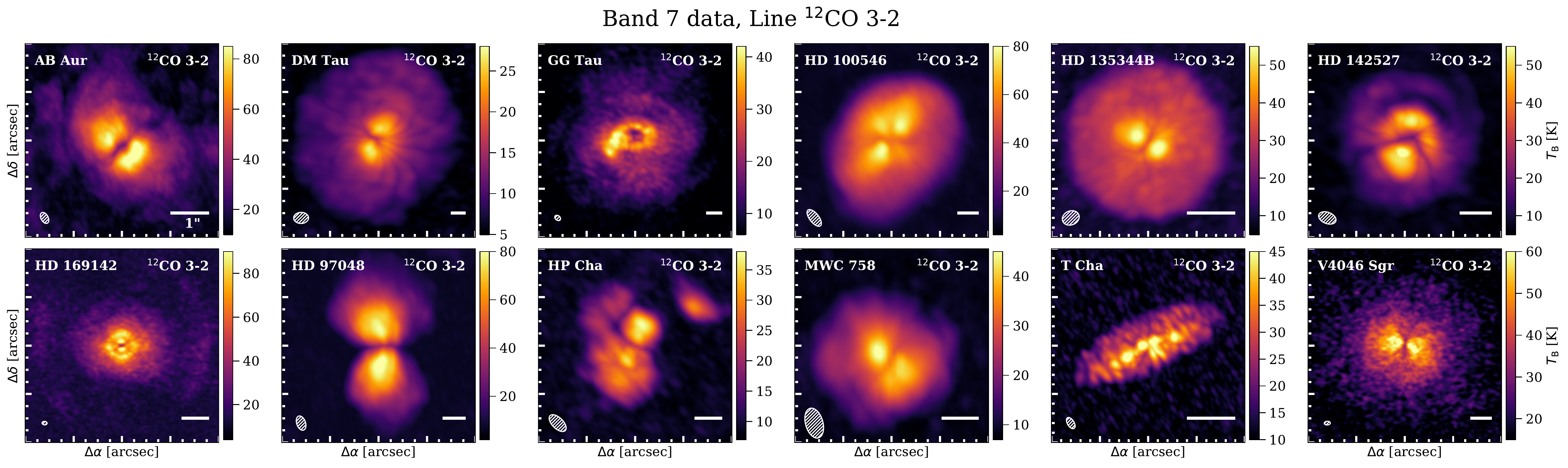}
\includegraphics[width=1.0\textwidth]{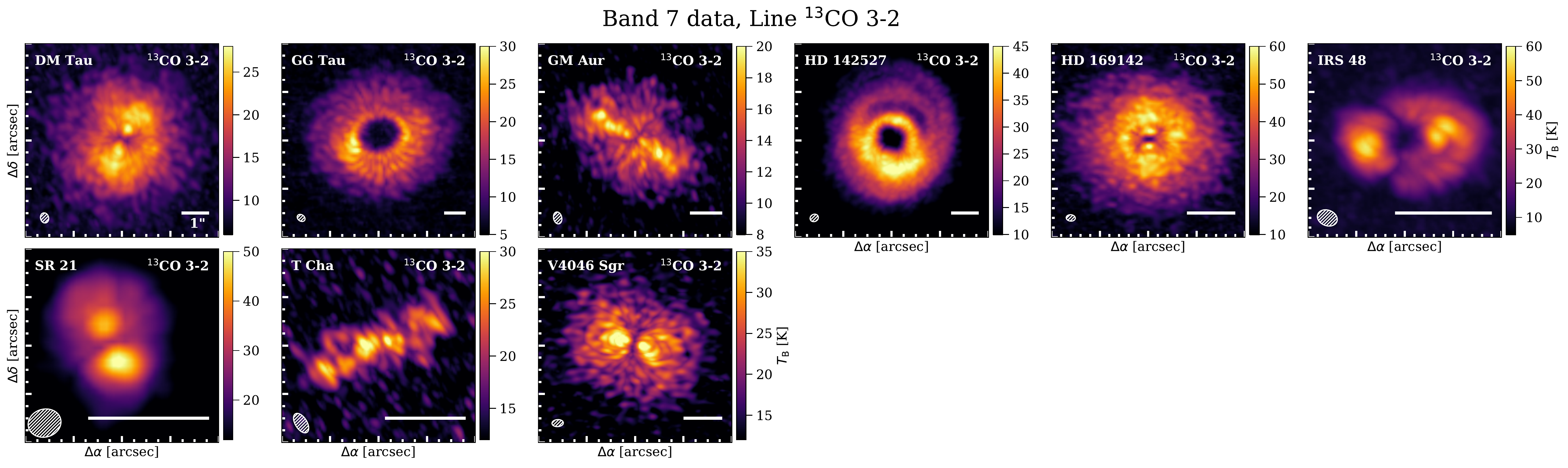}
\includegraphics[width=1.0\textwidth]{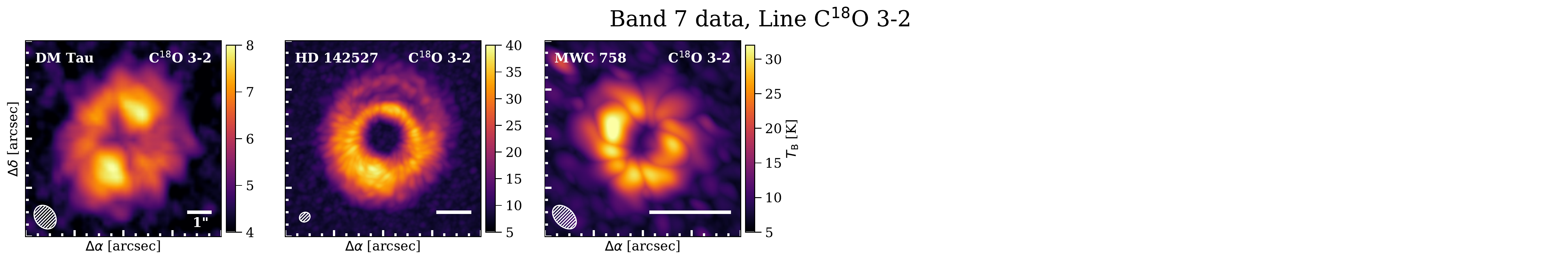}
\caption{Peak brightness temperature maps of the gas emission in our targets, shown for the additional Band 7 CO lines used in this analysis. The conversion from peak intensity to units of Kelvin is done with the Planck law. The circle and bar in the bottom left and bottom right corner of each panel indicate the beam and a $1\arcsec$ scale respectively.}\label{fig:TbAddBand7}
\end{figure*}

\hyperref[fig:TbAddBand6]{Figure~\ref*{fig:TbAddBand6}} and \hyperref[fig:TbAddBand7]{Fig.~\ref*{fig:TbAddBand7}} show the brightness temperature maps for the additional CO lines in Band 6 and Band 7 respectively. 
\newpage
\newpage
\subsection{Rotation velocity maps}\label{appendix:v0maps}
\begin{figure*}[h!]
\centering
\includegraphics[width=1.0\textwidth]{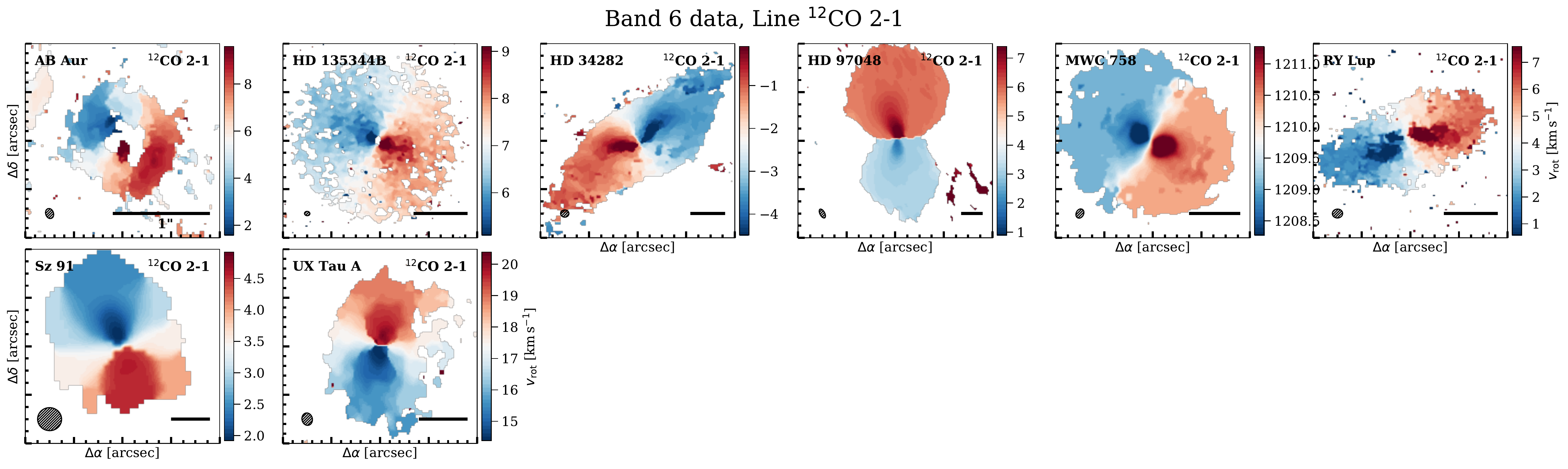}
\includegraphics[width=1.0\textwidth]{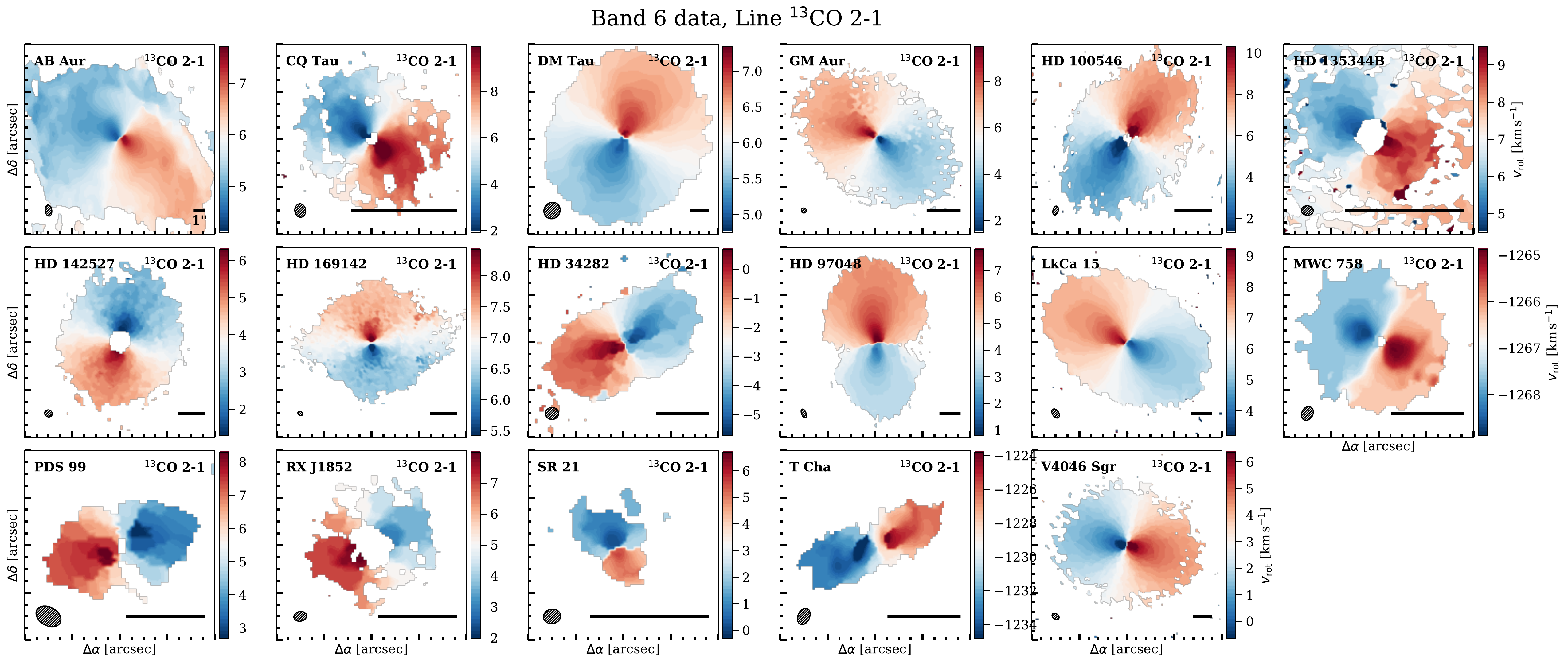}
\includegraphics[width=1.0\textwidth]{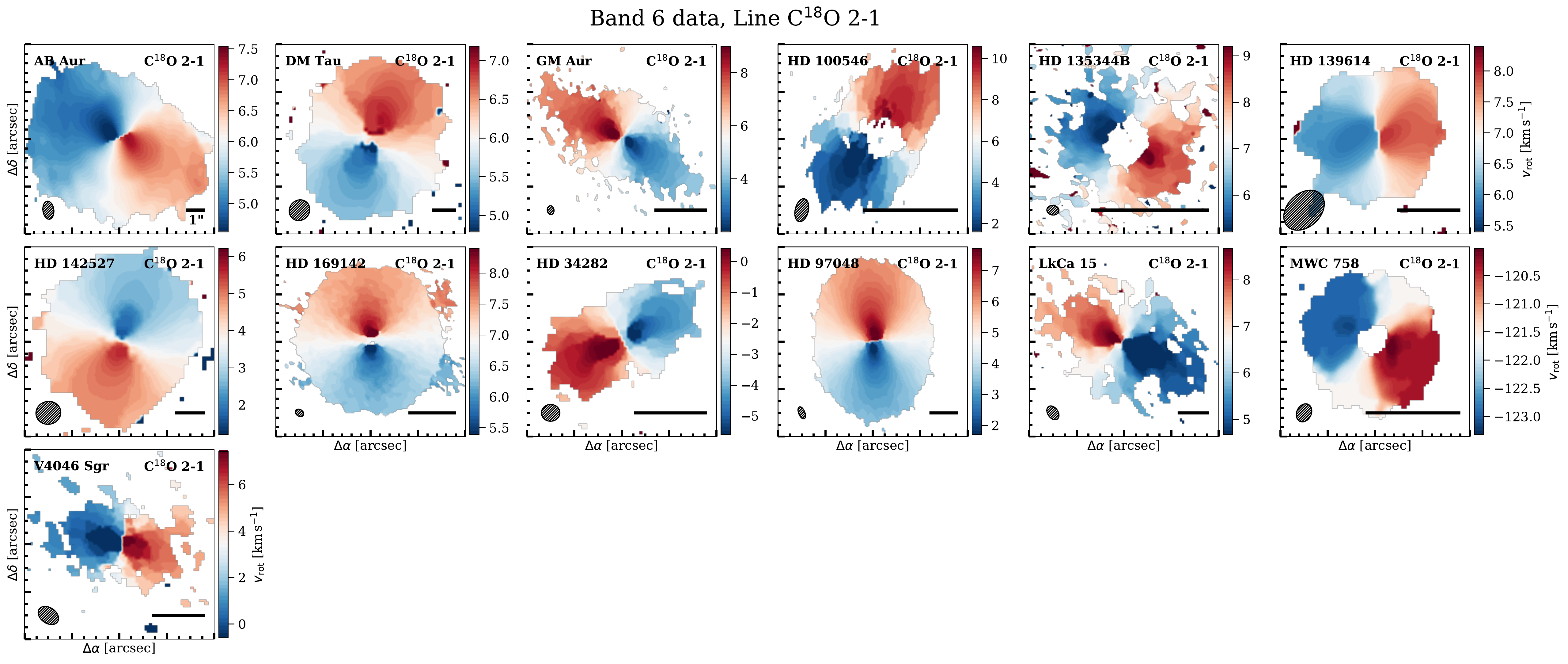}
\caption{Rotation velocity maps of the gas emission in our targets, shown for the additional Band 6 lines used in this analysis. The maps are computed with \texttt{bettermoments}. The circle and bar in the bottom left and bottom right corner of each panel indicate the beam and a $1\arcsec$ scale respectively.}\label{fig:v0AddBand6}
\end{figure*}

\begin{figure*}[h!]
\centering
\includegraphics[width=1.0\textwidth]{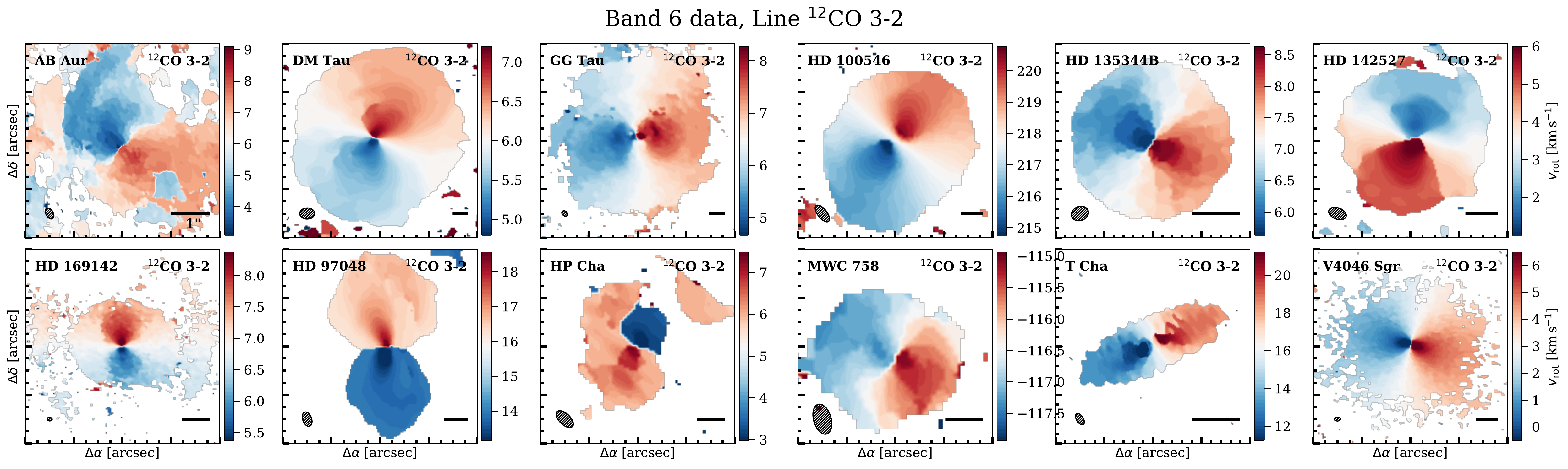}
\includegraphics[width=1.0\textwidth]{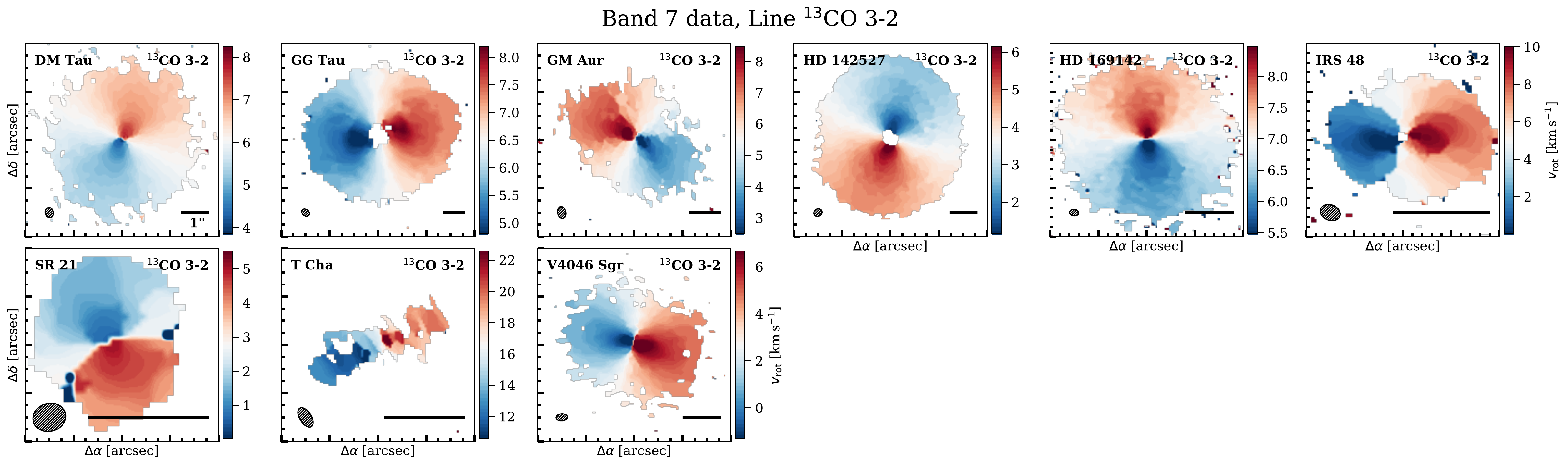}
\includegraphics[width=1.0\textwidth]{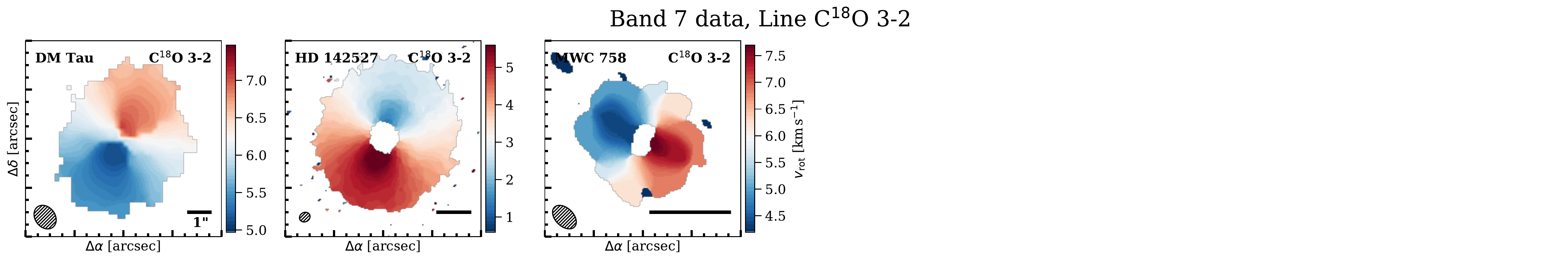}
\caption{Rotation velocity maps of the gas emission in our targets, shown for the additional Band 7 lines used in this analysis. The maps are computed with \texttt{bettermoments}. The circle and bar in the bottom left and bottom right corner of each panel indicate the beam and a $1\arcsec$ scale respectively.}\label{fig:v0AddBand7}
\end{figure*}

\hyperref[fig:TbAddBand6]{Figure~\ref*{fig:v0AddBand6}} and \hyperref[fig:TbAddBand7]{Fig.~\ref*{fig:v0AddBand7}} show the kinematical maps for the additional CO lines in Band 6 and Band 7 respectively. 
\newpage
\newpage
\subsection{Brightness temperature residuals}\label{appendix:tbresadd}
\begin{figure*}[h!]
\centering
\includegraphics[width=1.0\textwidth]{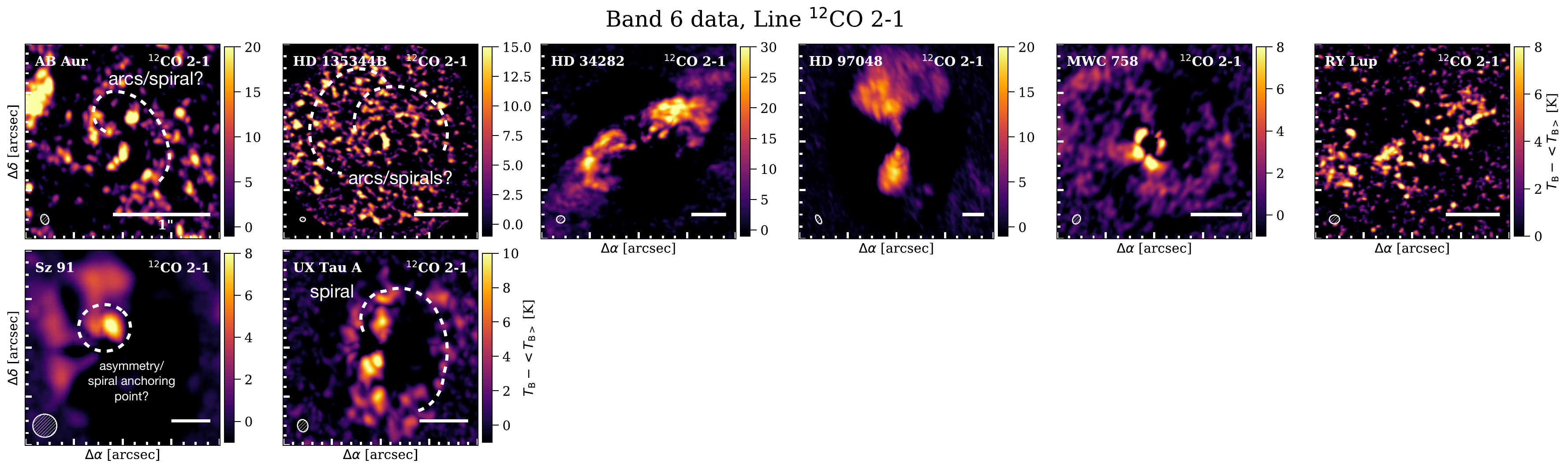}
\includegraphics[width=1.0\textwidth]{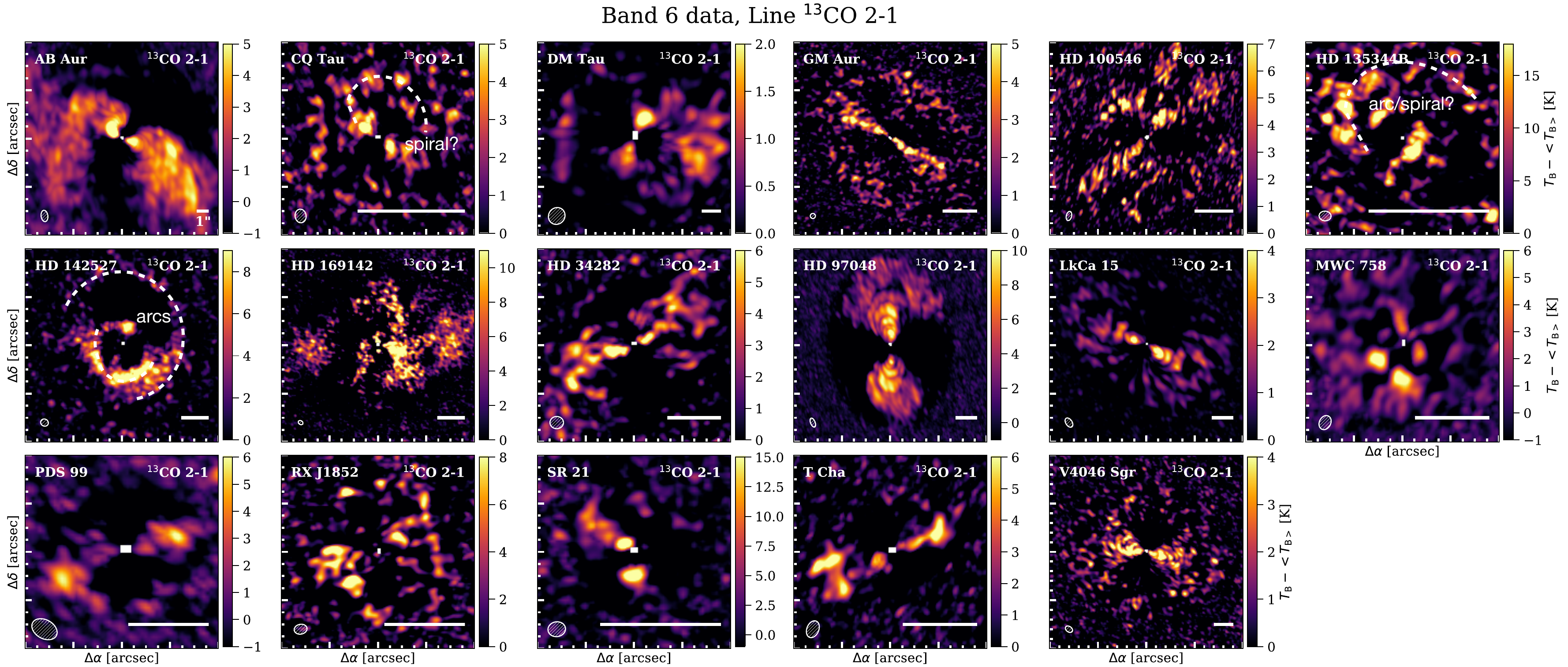}
\includegraphics[width=1.0\textwidth]{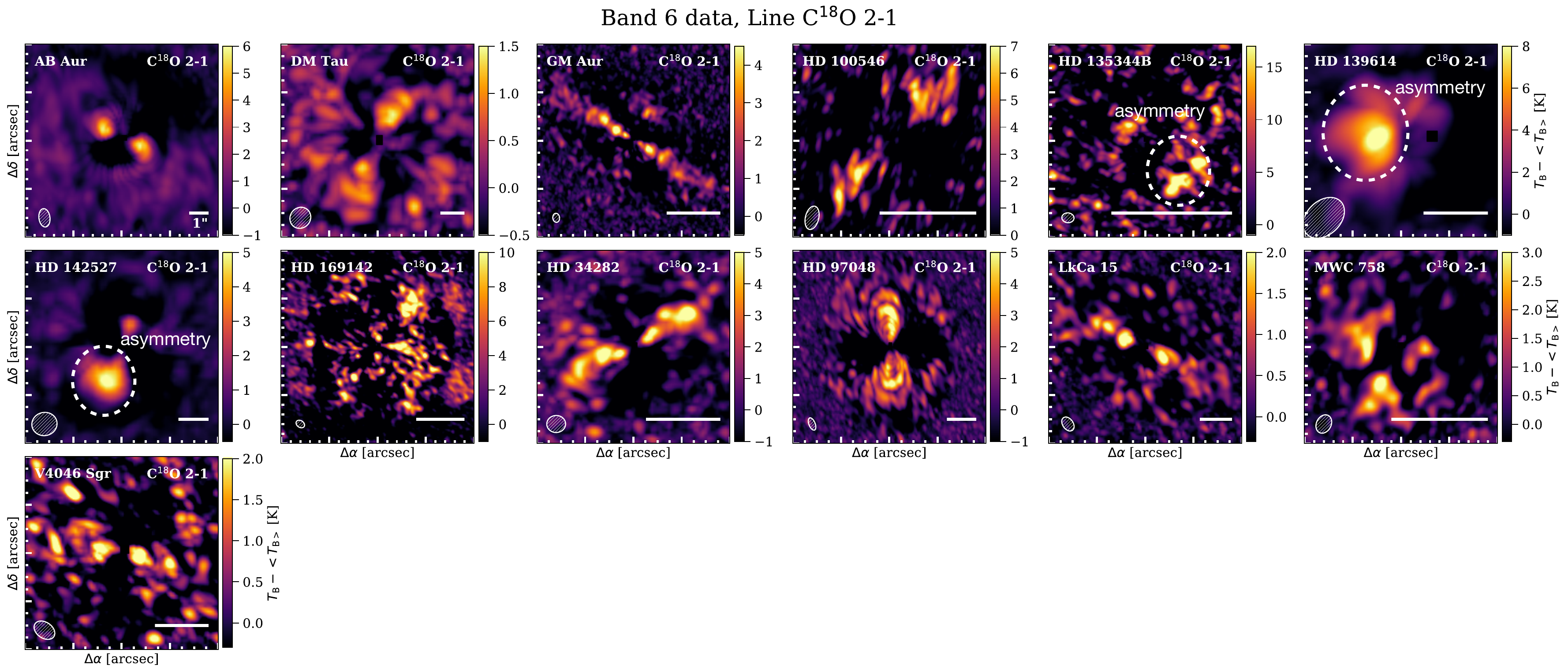}
\caption{Brightness temperature residuals shown for the additional Band 6 lines used in this analysis and for a thin disc geometry. Obtained with \texttt{GoFish}. The circle and bar in the bottom left and bottom right corner of each panel indicate the beam and a $1\arcsec$ scale respectively. Some features are annotated.}\label{fig:TbresB6thin}
\end{figure*}

\begin{figure*}[h!]
\centering
\includegraphics[width=1.0\textwidth]{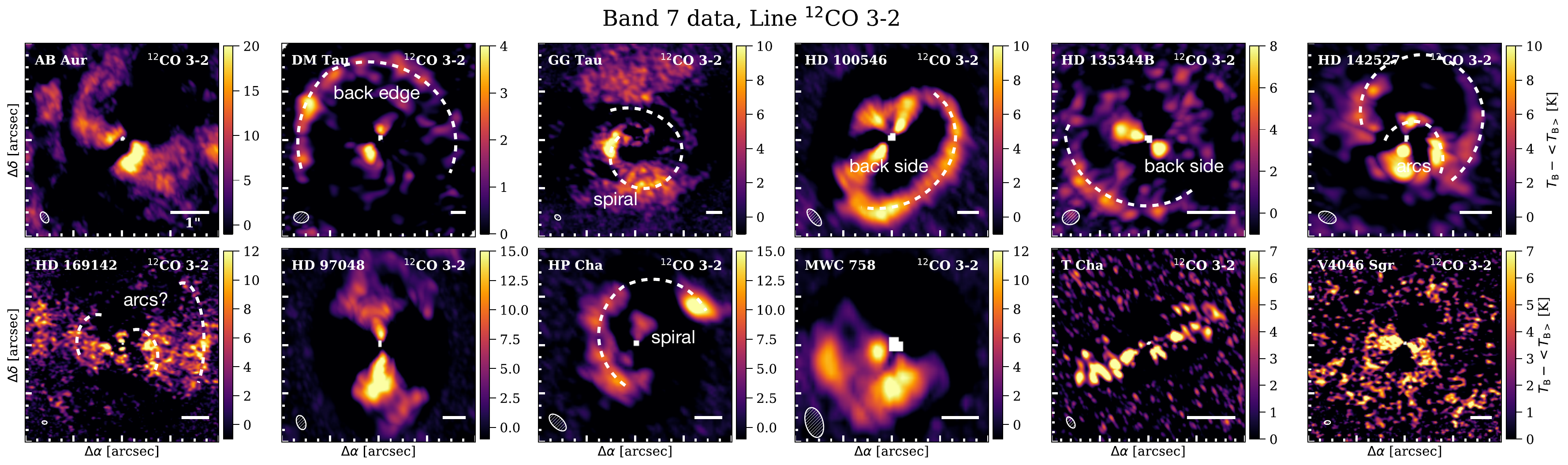}
\includegraphics[width=1.0\textwidth]{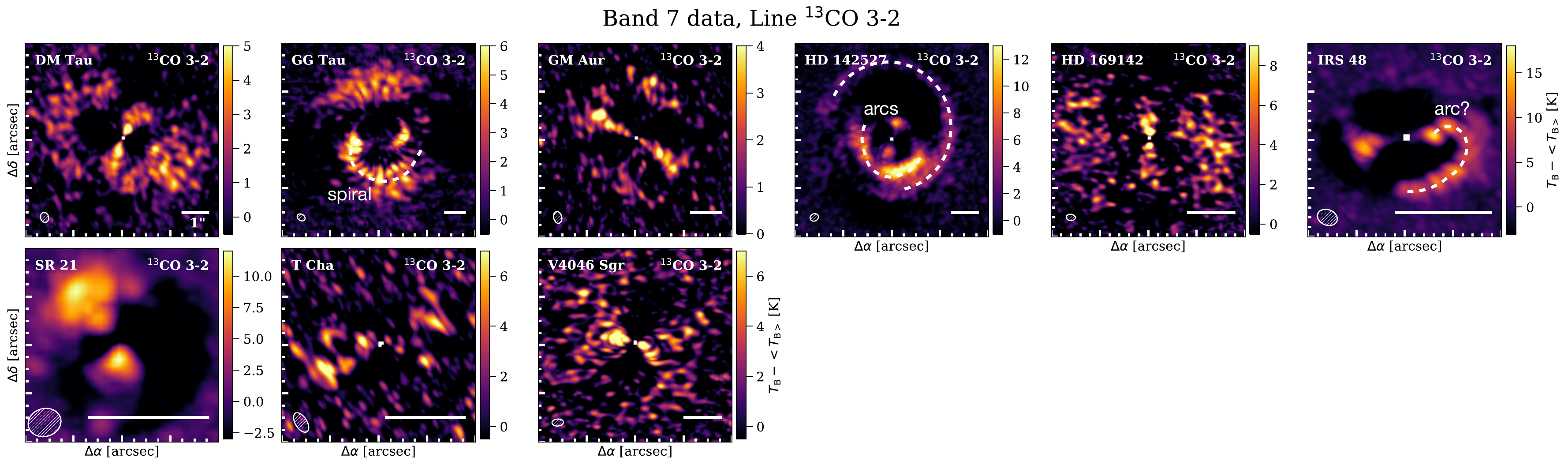}
\includegraphics[width=1.0\textwidth]{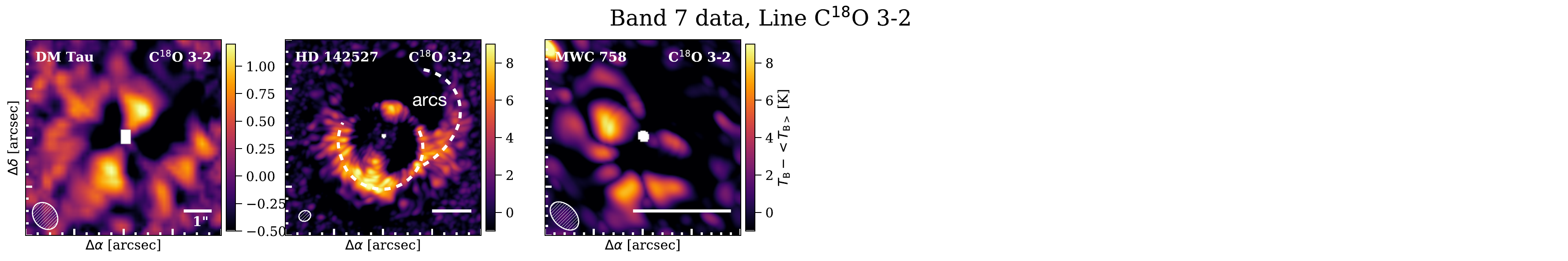}
\caption{Brightness temperature residuals shown for the additional Band 7 lines used in this analysis and for a thin disc geometry. Obtained with \texttt{GoFish}. The circle and bar in the bottom left and bottom right corner of each panel indicate the beam and a $1\arcsec$ scale respectively. Some features are annotated.}\label{fig:TbresB7thin}
\end{figure*}

\begin{figure*}[h!]
\centering
\includegraphics[width=1.0\textwidth]{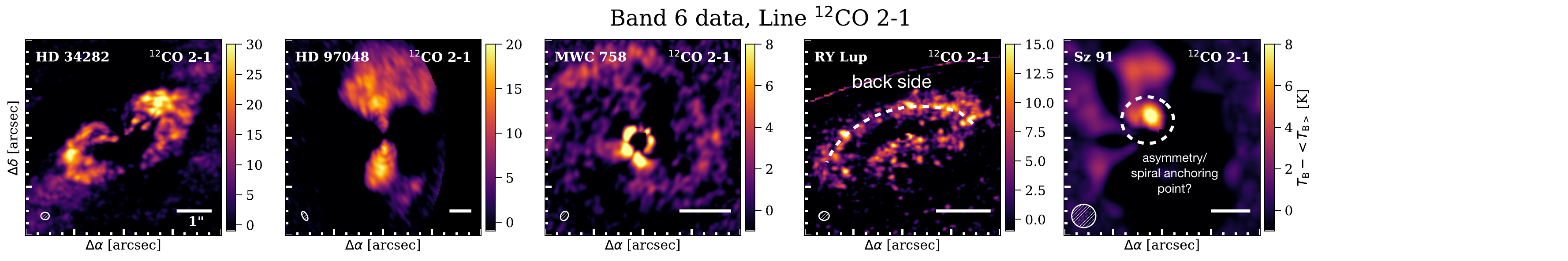}
\includegraphics[width=1.0\textwidth]{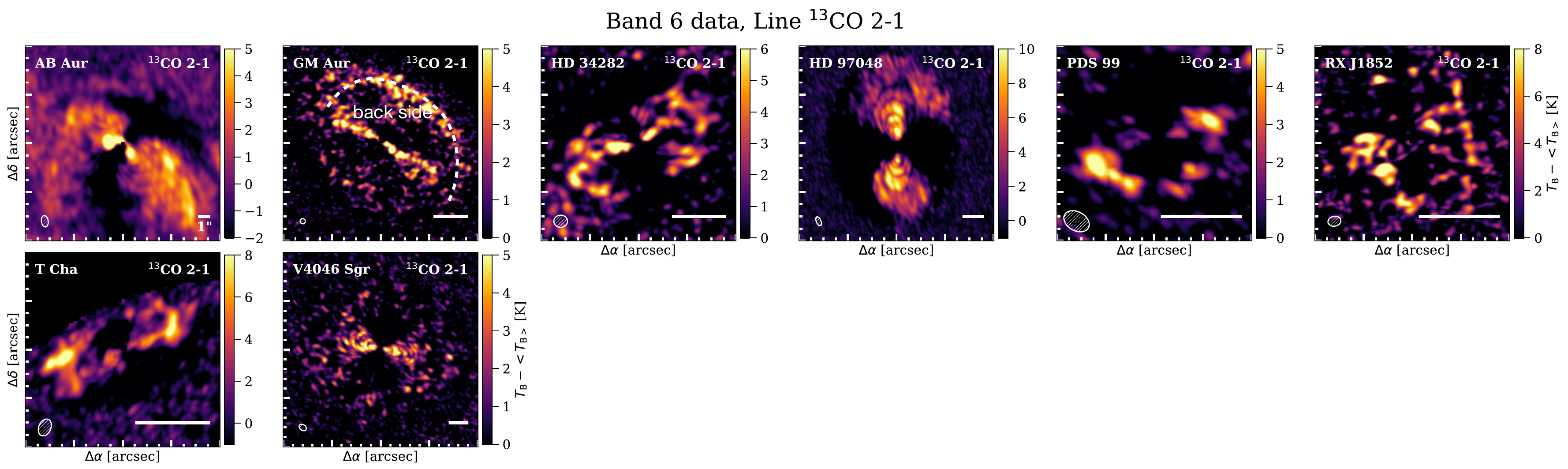}
\includegraphics[width=1.0\textwidth]{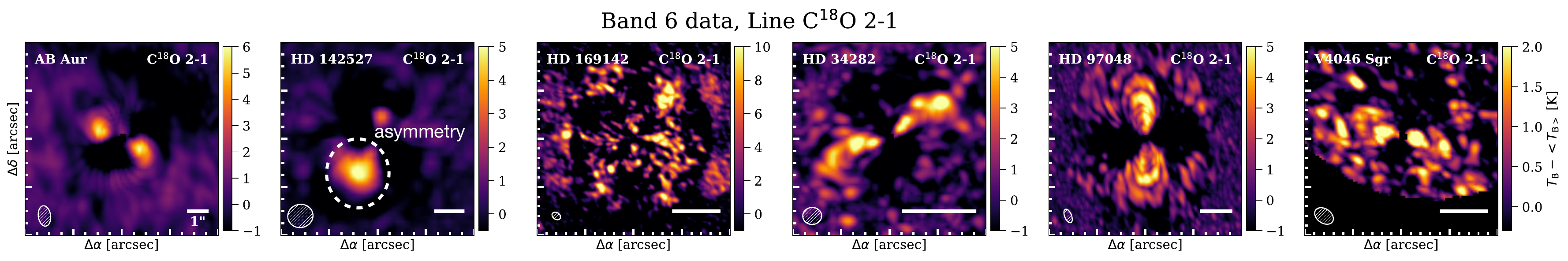}
\includegraphics[width=1.0\textwidth]{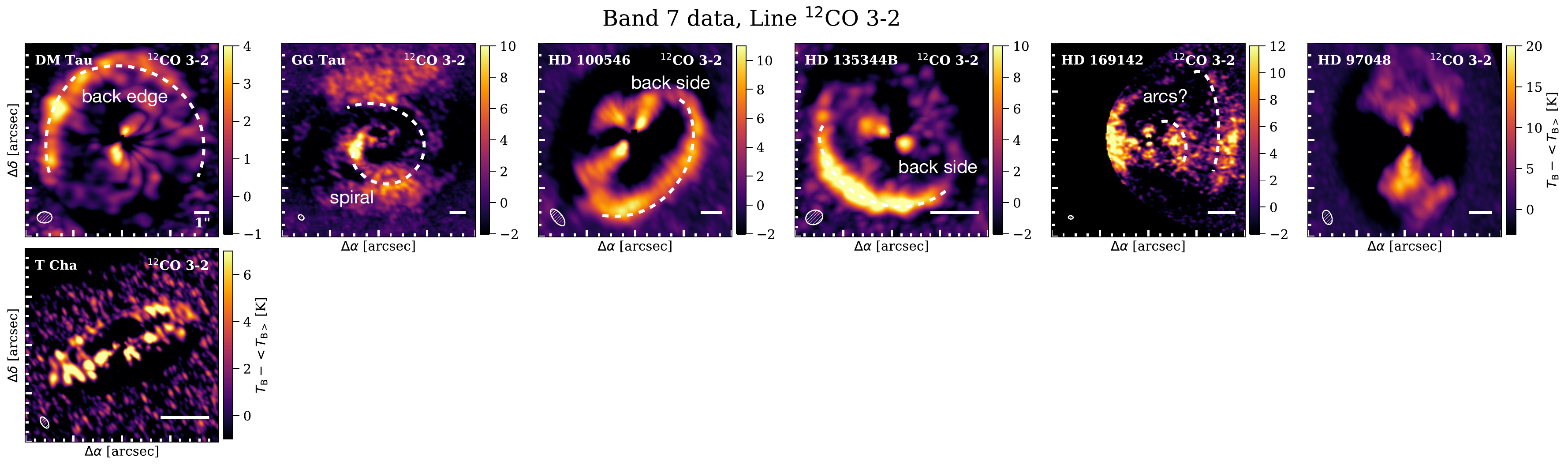}
\includegraphics[width=1.0\textwidth]{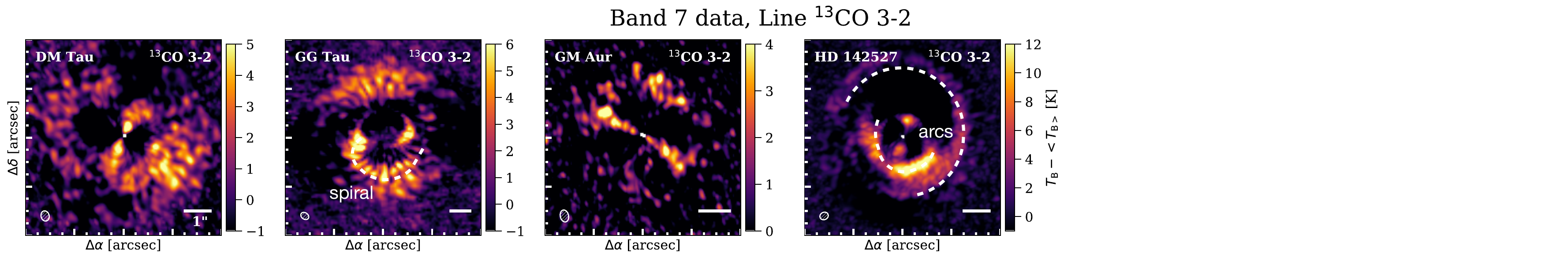}
\caption{Brightness temperature residuals shown for the additional Band 6 and Band 7 lines used in this analysis and for a thick disc geometry. Obtained with \texttt{GoFish}. The circle and bar in the bottom left and bottom right corner of each panel indicate the beam and a $1\arcsec$ scale respectively. Some features are annotated.}\label{fig:TbresB6B7thick}
\end{figure*}

\hyperref[fig:TbresB6thin]{Figure~\ref*{fig:TbresB6thin}}, \hyperref[fig:TbresB7thin]{Fig.~\ref*{fig:TbresB7thin}} and \hyperref[fig:TbresB6B7thick]{Fig.~\ref*{fig:TbresB6B7thick}} show the brightness temperature residuals for the additional CO lines in Band 6 and Band 7 for both the thin and the thick disc geometry.
\newpage
\newpage
\newpage
\subsection{Rotation velocity residuals}\label{appendix:v0resadd}
\begin{figure*}[h!]
\centering
\includegraphics[width=1.0\textwidth]{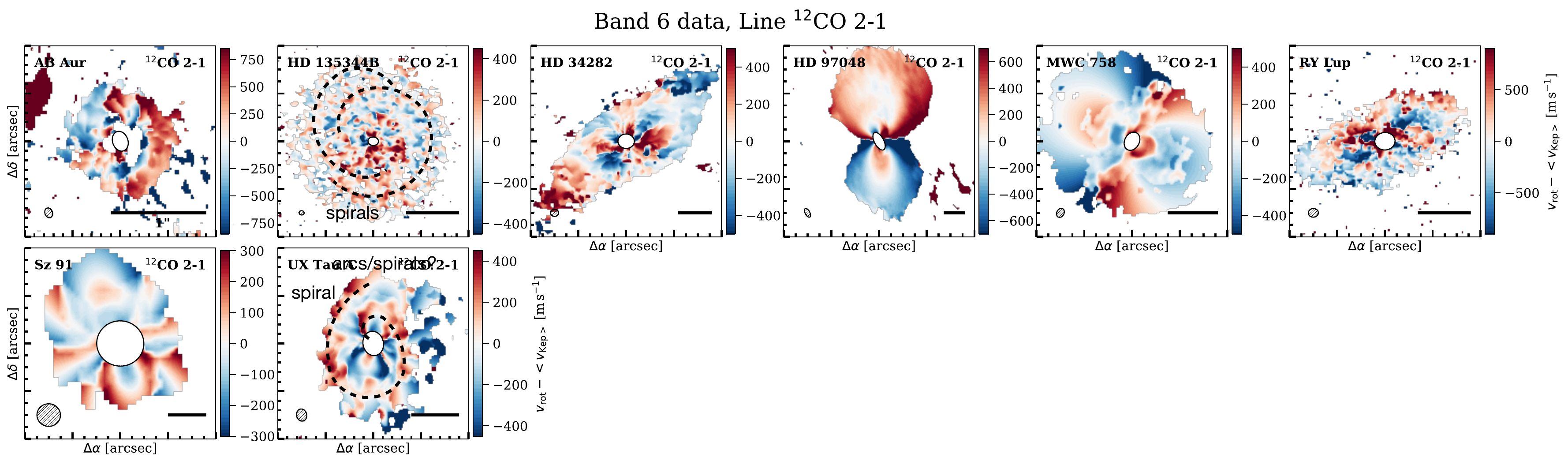}
\includegraphics[width=1.0\textwidth]{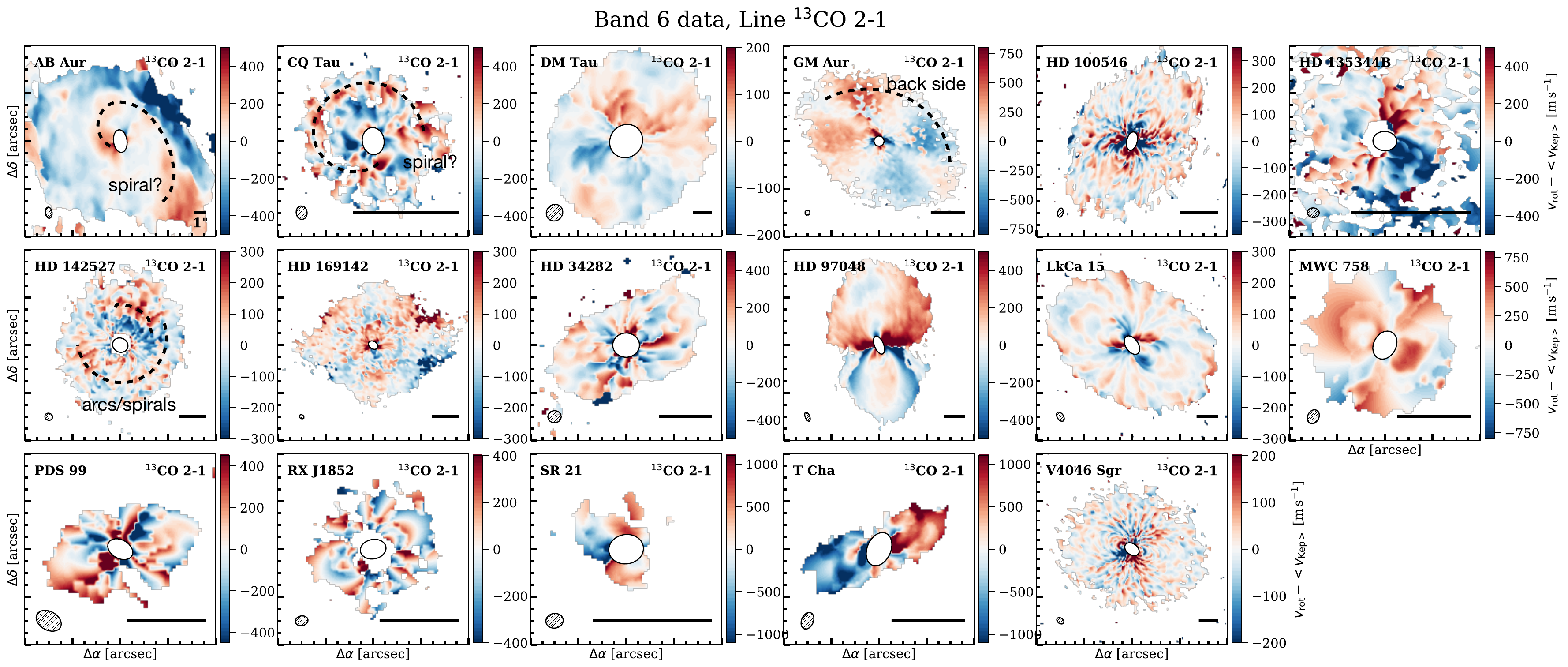}
\includegraphics[width=1.0\textwidth]{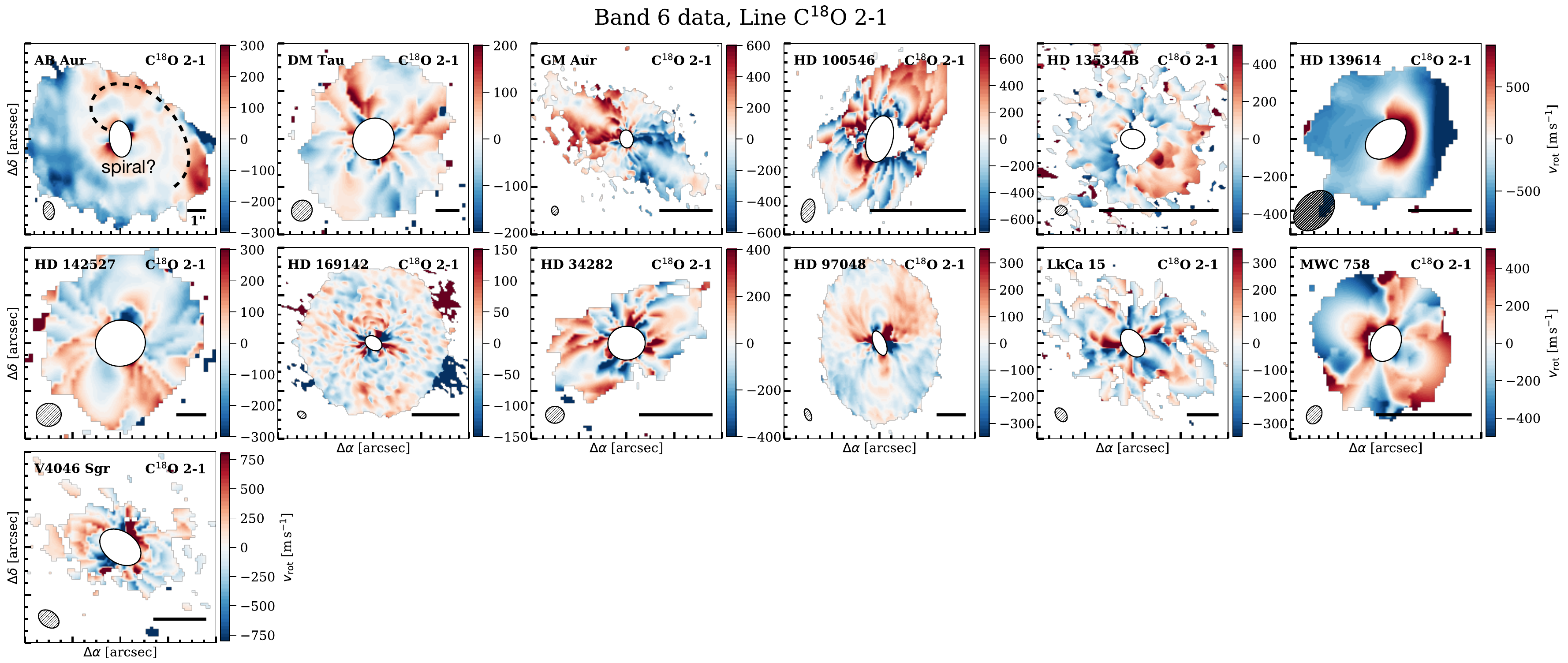}
\caption{Rotation velocity residuals shown for the additional Band 6 lines used in this analysis and for a thin disc geometry. Obtained with \texttt{eddy}. The circle and bar in the bottom left and bottom right corner of each panel indicate the beam and a $1\arcsec$ scale respectively. Some features are annotated.}\label{fig:v0resB6thin}
\end{figure*}

\begin{figure*}[h!]
\centering
\includegraphics[width=1.0\textwidth]{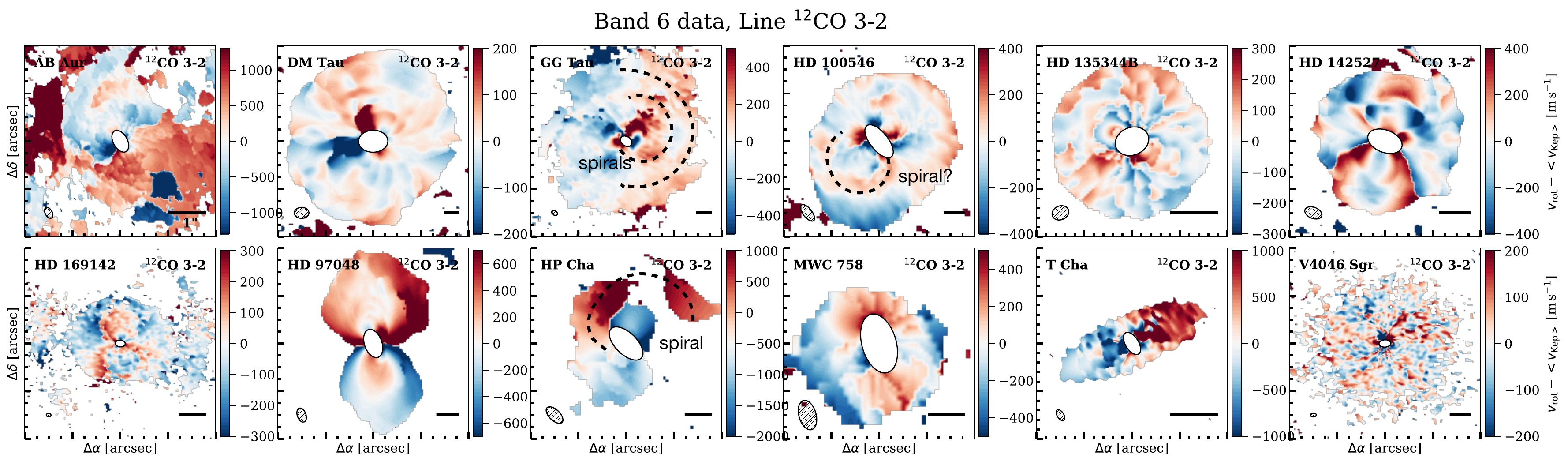}
\includegraphics[width=1.0\textwidth]{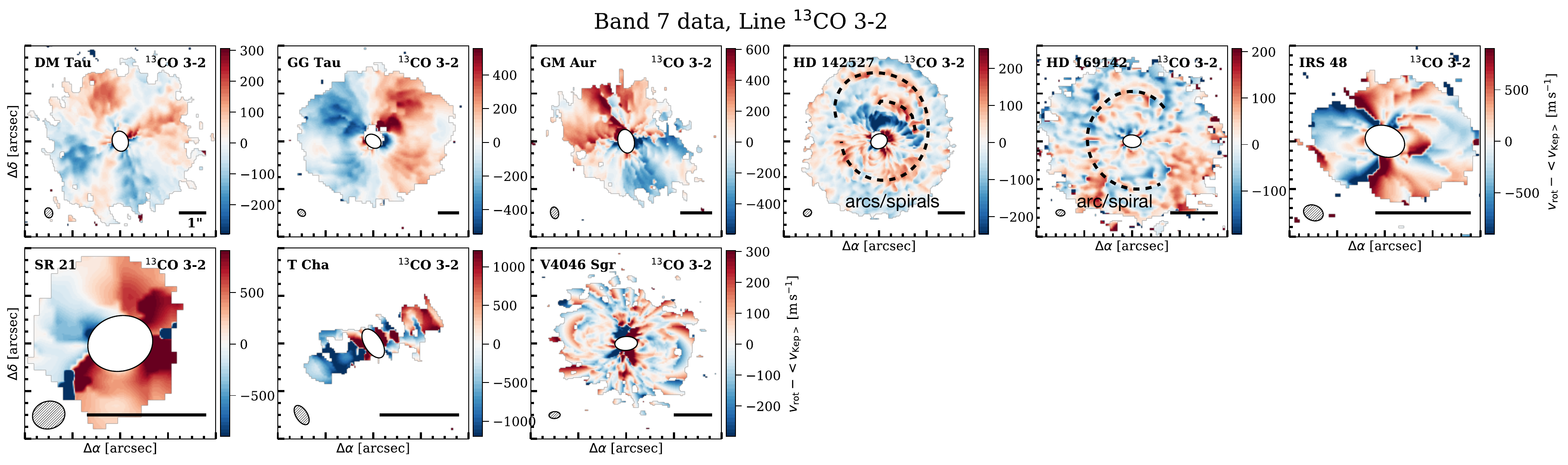}
\includegraphics[width=1.0\textwidth]{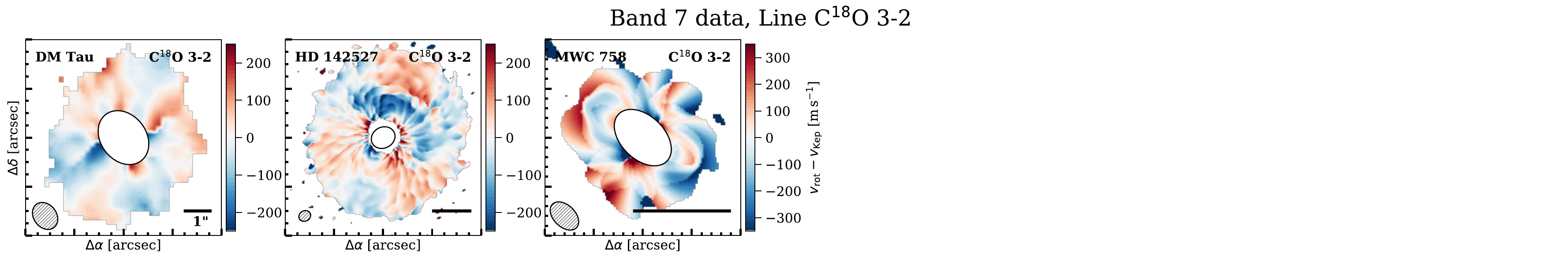}
\caption{Rotation velocity residuals shown for the additional Band 7 lines used in this analysis and for a thin disc geometry. Obtained with \texttt{eddy}. The circle and bar in the bottom left and bottom right corner of each panel indicate the beam and a $1\arcsec$ scale respectively. Some features are annotated.}\label{fig:v0resB7thin}
\end{figure*}

\begin{figure*}[h!]
\centering
\includegraphics[width=1.0\textwidth]{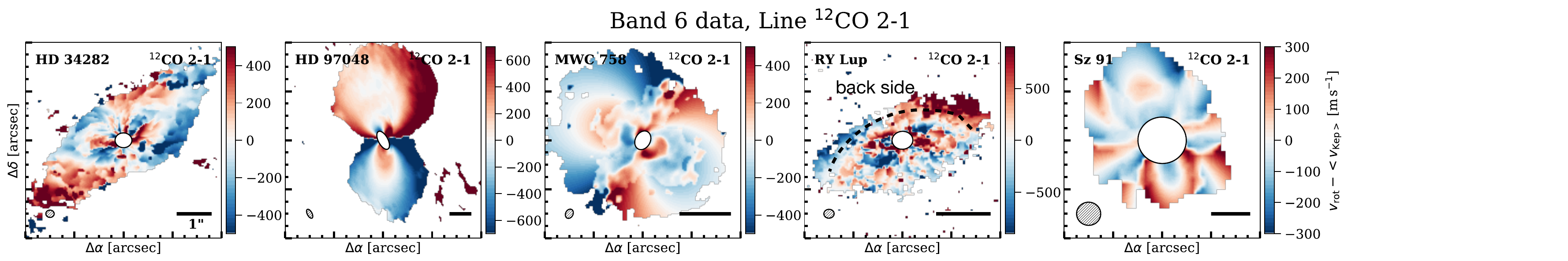}
\includegraphics[width=1.0\textwidth]{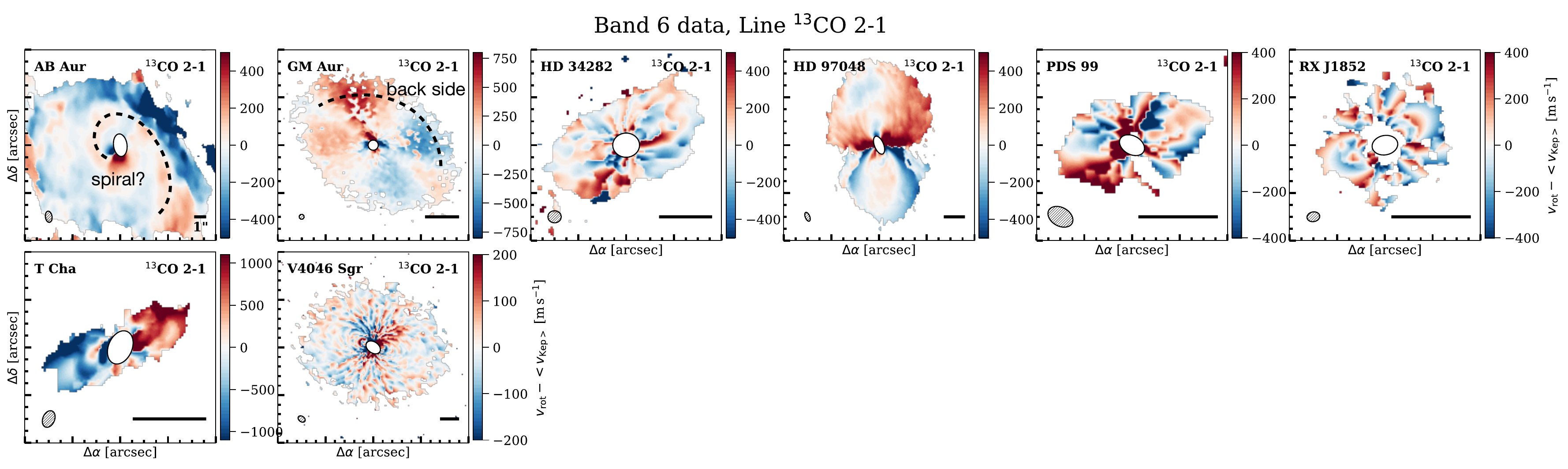}
\includegraphics[width=1.0\textwidth]{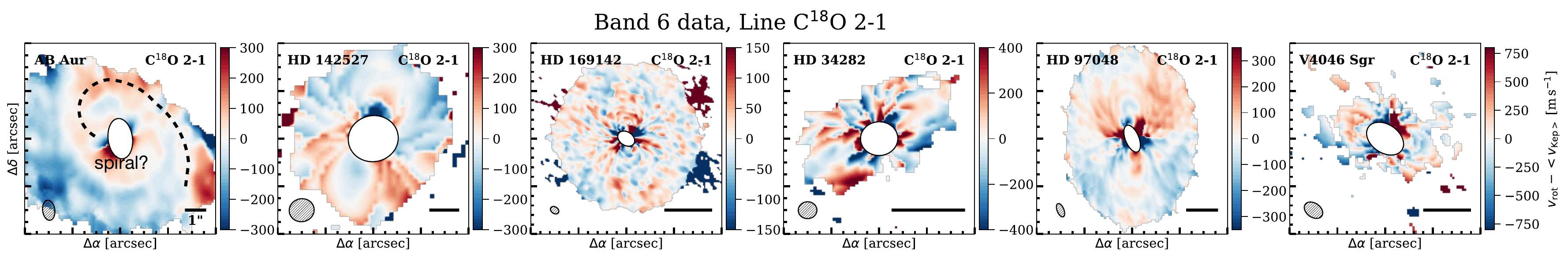}
\includegraphics[width=1.0\textwidth]{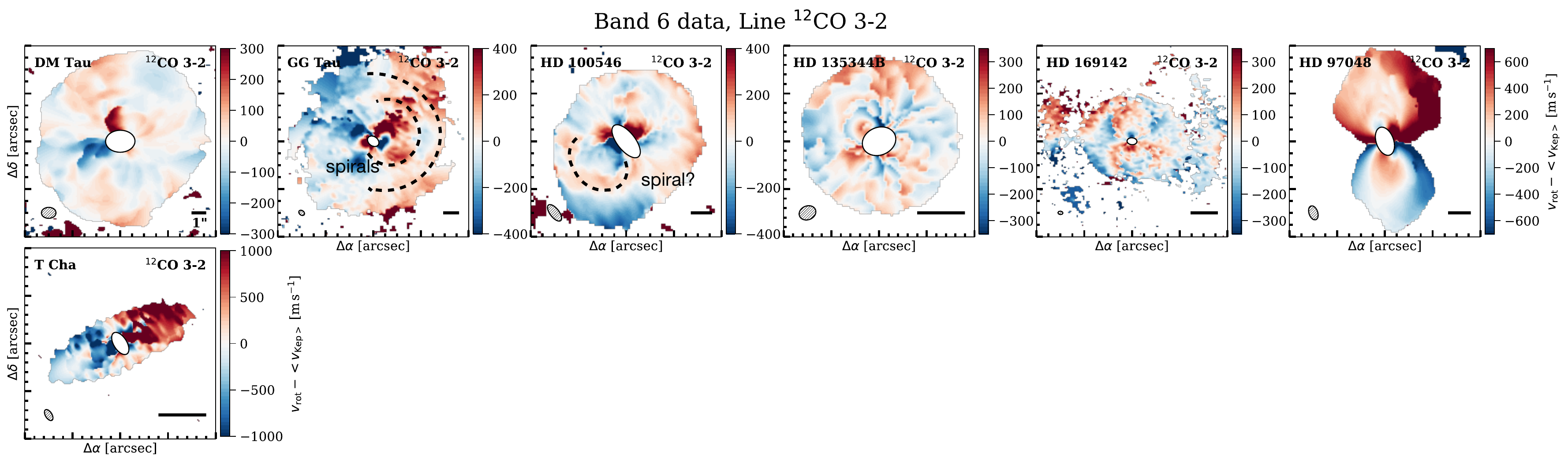}
\includegraphics[width=1.0\textwidth]{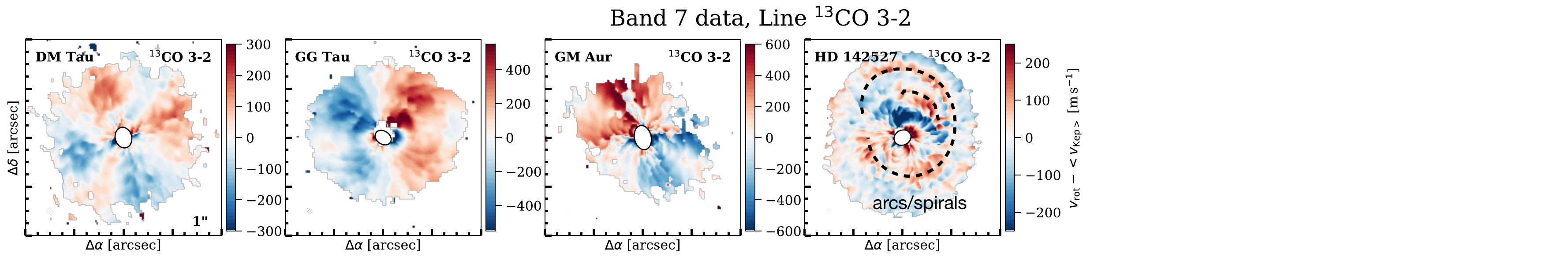}
\caption{Rotation velocity residuals shown for the additional Band 6 and Band 7 lines lines used in this analysis and for a thick disc geometry. Obtained with \texttt{eddy}. The circle and bar in the bottom left and bottom right corner of each panel indicate the beam and a $1\arcsec$ scale respectively. Some features are annotated.}\label{fig:v0resB6B7thick}
\end{figure*}

\hyperref[fig:v0resB6thin]{Figure~\ref*{fig:v0resB6thin}}, \hyperref[fig:v0resB7thin]{Fig.~\ref*{fig:v0resB7thin}} and \hyperref[fig:v0resB6B7thick]{Fig.~\ref*{fig:v0resB6B7thick}} show the rotation velocity residuals for the additional CO lines in Band 6 and Band 7 for both the thin and the thick disc geometry.
\newpage
\newpage
\newpage
\section{Radial profiles of the integrated intesity}
\begin{figure*}[h!]
\centering
\includegraphics[angle=90,width=0.86\textwidth]{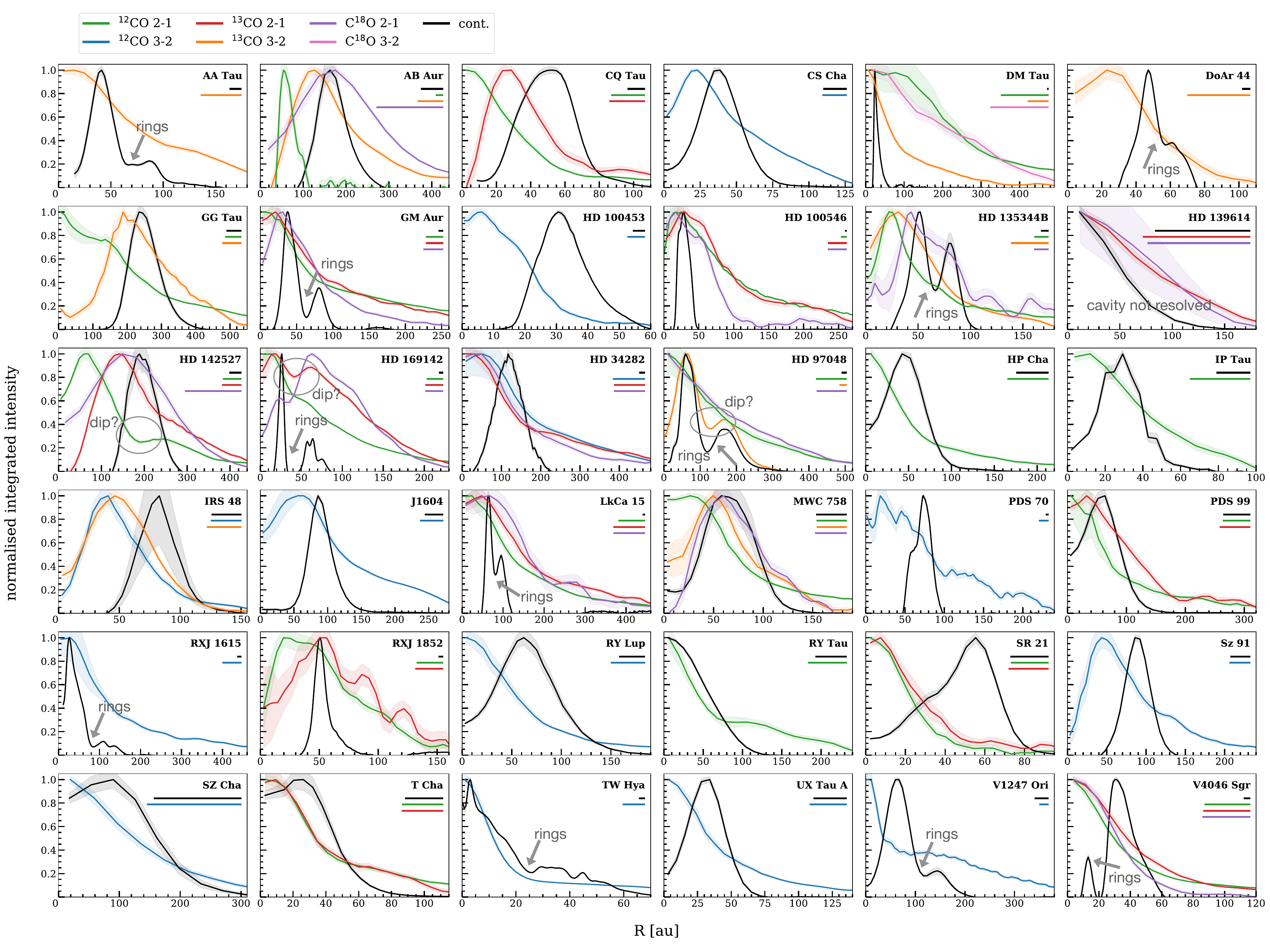}
\caption{Azimuthally averaged and normalised radial integrated intensity profiles for the different CO lines (colored lines) and continuum (black lines) emission. The major beam of each observations is indicated by the bars in the top right corner. Some features of the profiles are annotated in the individual panels.}\label{fig:radialProfiles2}
\end{figure*}

In \hyperref[fig:radialProfiles2]{Fig.~\ref*{fig:radialProfiles2}} the radial profiles of the integrated intensity are shown for different CO isotopologues. Such profiles can be used to estimate the size of the cavity. As a deep cavity we define those cases where a clear drop of the emission can been seen in the inner regions for at least the more optically thin lines that tend to trace the column density.
%
%
%
\end{appendix}
\end{document}